\documentclass[fleqn,usenatbib]{mnras}

\usepackage{newtxtext,newtxmath}

\usepackage[T1]{fontenc}

\DeclareRobustCommand{\VAN}[3]{#2}
\let\VANthebibliography\thebibliography
\def\thebibliography{\DeclareRobustCommand{\VAN}[3]{##3}\VANthebibliography}

\usepackage{graphicx}	\usepackage{amsmath}	
	\usepackage[dvipsnames]{xcolor}
\usepackage{multirow}

\newcommand{\redit}[1]{{#1}}
\newcommand{\rredit}[1]{{#1}}
\newcommand{\rrredit}[1]{{#1}}

\title[Constraining the $z\sim6$ XLF with ExSeSS]{Constraints on the X-ray Luminosity Function of AGN at $z=5.7-6.4$ with the Extragalactic Serendipitous Swift Survey}

\author[C. L. Barlow-Hall et al.]{
C. L. Barlow-Hall,$^{1}$\thanks{E-mail: c.barlow-hall@roe.ac.uk}
J. Delaney,$^{1}$
J. Aird,$^{1,2}$
P. A. Evans,$^2$
J. P. Osborne,$^2$
and
M. G. Watson$^2$
\\
$^{1}$Institute for Astronomy, The University of Edinburgh, Royal Observatory, Blackford Hill, Edinburgh EH9 3HJ, UK\\
$^2$School of Physics \& Astronomy, University of Leicester, University Road, Leicester LE1 7RJ, UK\\
}

\date{Submitted to MNRAS 2022-Jan-25}

\pubyear{2022}

\begin{document}
\label{firstpage}
\pagerange{\pageref{firstpage}--\pageref{lastpage}}
\maketitle

\begin{abstract}

X-ray luminosity functions (XLFs) of Active Galactic Nuclei (AGN) trace the growth and evolution of supermassive black hole populations across cosmic time, however, current XLF models are poorly constrained at redshifts of $z>6$, \rrredit{with a lack of spectroscopic constraints at these high redshifts}. In this work we \redit{place limits} on the bright-end of the XLF at $z=5.7-6.4$ using high-redshift AGN identified within the Extragalactic Serendipitous Swift Survey (ExSeSS) catalogue. Within ExSeSS we find one \rredit{serendipitously X-ray detected $z>6$ AGN}, ATLAS J025.6821-33.4627, with an X-ray luminosity of $L_\mathrm{X}=8.47^{+3.40}_{-3.13}\times10^{44}\mathrm{erg.s^{-1}}$ and $z=6.31\pm0.03$ making it the highest redshift, spectroscopically confirmed, serendipitously \rredit{X-ray detected} quasar known to date. 
We also calculate an upper limit on the space density at higher luminosities where no additional sources are found, enabling us to place constraints on the shape of the XLF.
Our results are consistent with the rapid decline in the space densities of high-luminosity AGN toward high redshift as predicted by extrapolations of existing parametric models of the XLF. 
We also find that our X-ray based measurements are consistent with estimates of the bolometric quasar luminosity function based on UV measurements at $z\gtrsim6$, although they require a large X-ray to bolometric correction factor at these high luminosities.
\end{abstract}

\begin{keywords}
galaxies: active – galaxies: evolution – galaxies: luminosity function, mass function – X-rays: galaxies.
\end{keywords}

\section{Introduction}\label{sec:Intro}

Most galaxies are thought to play host to a Supermassive Black Hole (SMBHs), with SMBHs and galaxies thought to co-evolve \citep{Kormendy&Ho2013}. When rapidly growing these SMBHs produce strong emissions across a range of wavelengths, from radio to high-energy X-rays, powered by their accretion activity \citep[e.g. see reviews of][]{Reines&Comastri2016, Padovani2017, Hickox&Alexander2018}. These luminous \rrredit{systems} are known as Active Galactic Nuclei (AGN) and form the basis for investigations of SMBHs beyond our local Universe, with AGN observed throughout the Universe and even out at redshifts of $z>6$.

Large-scale optical and near-infrared imaging surveys have enabled the identification of luminous AGN out to $z=7.54$ \citep{Banados2018} and $z=7.642$ \citep[][]{Wang2021}. Spectroscopic follow-up observations not only confirm the redshifts of these sources but also reveal them to have remarkably similar rest-frame UV spectra to their lower redshift counterparts \citep[e.g.][]{Mortlock2011, DeRosa2014, Shen2019}. 
Applying single epoch scaling relations indicates that they have masses of $\sim10^{6-8}\mathrm{M_{\sun}}$ \citep[\rrredit{see e.g.}][\rrredit{and references therein}]{Onoue2019,Zubovas&King2021,Yang2021}, which are comparable to SMBH masses in the nearby Universe. This raises the question as to how these SMBHs formed and grew to these masses within the short time period of the early Universe. 

The main seeding mechanisms theorised for the formation of AGN are Population III stellar remnants \citep[e.g.][]{Madau&Rees2001} and Direct Collapse Black Holes \citep[e.g.][]{Volonteri2010}, producing Black Hole seeds of masses $10-100\mathrm{M_{\odot}}$ or $10^4-10^6\mathrm{M_{\odot}}$ respectively. Thus, even for the case of direct collapse, a significant amount of growth must have occurred within the first few 100~Myrs of cosmic time in order for these seed black holes to have attained the masses we observe. However, this growth remains poorly constrained, due to the lack of robust observational constraints on AGN within the early Universe.

The growth of AGN populations across cosmic time and the evolution of AGN accretion rate is traced by the Quasar Luminosity Function (QLF). The QLF describes the comoving space density of AGN as a function of redshift and luminosity \citep[e.g.][]{Boyle2000, Page1996, Kalfountzou2014} and is measured using surveys of AGN selected using optical, IR and X-ray data \citep{Hopkins2007, Ross2013}. Many AGN have been discovered through rest-frame optical/UV selection \citep[e.g.][]{Banados2016}, which is probed by optical and IR surveys that cover large areas of sky identifying  AGN out to very high redshifts \citep[\rrredit{e.g.}][]{McGreer2013,Matsuoka2019,Reed2019,Wang2019}. X-ray follow-up of high-redshift AGN samples identified through optical or IR surveys enable further investigations the nature of these sources \citep{Brandt2002, Vignali2001, Vito2019}. However, rest-frame optical/UV selection is biased towards the most luminous AGN sources, as these are more easily detectable by optical/UV telescopes above the emissions of the host galaxy. Processes within the host galaxies can also contaminate the AGN light at optical/UV and IR wavelengths, unlike X-ray selected AGN samples.

X-ray selection is often used to identify samples of AGN without the strong bias toward the most luminous sources that affects optical/UV selection. This lack of bias arises as few processes within galaxies produce significant X-ray emission and thus AGN easily outshine their host galaxies at X-ray wavelengths \citep[see e.g.][]{Padovani2017}. Furthermore, X-ray emission is much less susceptible to obscuration effects than the optical/UV light. 
Thus, AGN can be efficiently identified using X-ray surveys, with the accretion rate and \rredit{hence the} growth of the central SMBH being reliably traced by the X-ray emission. 
Thanks to their well-defined sensitivity and uncontaminated selection of AGN, X-ray surveys are especially useful for determining the QLF. The X-ray QLF, known as the X-ray Luminosity Function (XLF) can then be used to place constraints on the activity of AGN across cosmic time and thus the rate of growth of the early population of SMBHs. 

Prior studies of the XLF have shown that AGN populations evolve substantially over cosmic time, increasing in both space density and their typical luminosities between $z\sim0$ and $z\sim2$, where the overall accretion rate density peaks \citep[e.g.][]{Ueda2014,Aird2015}.
Toward higher redshifts ($z\gtrsim3$), the normalisation of the XLF is found to drop rapidly across all luminosities \citep[e.g.][]{Brusa2009,Vito2014,Georgakakis2015}, placing constraints on the extent of SMBH growth in the early Universe.
However, the samples of X-ray selected AGN at 
$z\gtrsim5$ remain extremely small: 
2 with photometric redshifts in the $\approx7\mathrm{Ms}$ Chandra Deep Field South \citep{Luo2017}, 2 with spectroscopic redshift (the highest at $z=5.3$) and 7 with photometric redshifts (4 of which are at $z>6$ with the highest at $z=6.85$) in the Chandra COSMOS-Legacy survey \citep[][]{Marchesi2016}. 
These small numbers  are due to both the strong decline in the XLF of AGN at high redshift, which can also be seen in the space density measured from optical/UV and IR surveys, and the depths required in order to detect even intrinsically luminous AGN at these extreme redshifts. Due to these small X-ray samples the parametric models of the XLF are poorly constrained at high redshift by current observations. 
Yet even with samples of just a few AGN at these very high redshifts, we can begin to place important constraints on the XLF. 

With the launch of eROSITA---providing a new generation of sensitive, wide-area X-ray surveys \citep{Predehl2021}---there is the potential to discover many more of the rare, high-luminosity X-ray selected AGN at $z\gtrsim6$ and improve our constraints on the XLF. 
Indeed, \citet{Khorunzhev2021} and \citet{Medvedev2020} have reported the discovery of highly luminous X-ray emission from quasars at $z\approx5.5$ and $z=6.18$, respectively, in the early all-sky eROSITA scans. \citet{Wolf2021} placed constraints on the XLF at $z\sim6$ using a single X-ray detected quasar at $z = 5.81$, found in the eROSITA Final Equatorial Depth Survey (eFEDS) that provides performance verification data  in a $\sim$140~deg$^2$ field at the depth of the final 4-year eROSITA all-sky surveys \citep{Brunner2021}.

In this paper, we present the \rredit{observational constraints on the XLF at $z>6$ given by sources} within the Extragalactic Serendipitous Swift Survey (ExSeSS) catalogue \citep{DelaneyPrep}.
ExSeSS covers a total sky area of $\sim$2000~deg$^2$ and reaches ultimate flux limits of $f_\mathrm{0.3-10keV}\sim10^{-15}$~erg~s$^{-1}$~cm$^{-2}$ for $\sim0.1$\% of the area, which are considerably deeper than the current \textit{eROSITA} all-sky coverage.  
We identify one X-ray source within ExSeSS that is associated with a previously known $z>6$ quasar with a spectroscopic redshift.
\rredit{Given the serendipitous nature of ExSeSS both the sensitivity and, consequently, the survey volume can be well-defined, allowing us}
to place direct observational constraints on the space density of luminous X-ray AGN at $z>6$. We then compare the estimated XLF and limits to extrapolated model XLFs from previous studies. 
The source catalogue used in this study, ExSeSS, is introduced in \S\ref{sec:ExSeSS}, while our process to identify high-redshift sources is outlined in \S\ref{sec:Sources}. In \S\ref{sec:XLFconstraints,sub:numbers} we compare predicted source yields based on extrapolations of current XLF models to our observed sample, while in \S\ref{sec:XLFconstraints,sub:XLF} we use our data to place constraints on the XLF. We also compare the constraints from our X-ray observations to existing estimates of the bolometric Quasar Luminosity Function from rest-frame optical/UV samples (\S\ref{sec:OpticalAndBol,sub:BolQLF}). We summarise our findings and present our conclusions in \S\ref{sec:Conclusions}. Throughout this paper we assume flat $\Lambda$CDM cosmology, with $H_0=70.0$, $\Omega_{\gamma}=0.7$ and $\Omega_{m}=0.3$, and all errors given are the $1\sigma$ uncertainties on the values.

\section{The ExSeSS catalogue}\label{sec:ExSeSS}

The studies performed in this paper make use of the sample of X-ray sources from the Extragalactic Serendipitous Swift Survey  (ExSeSS), as defined by \citet{DelaneyPrep}. 
Here, we give a brief overview of the sample construction and the vital step to define the area coverage and sensitivity of the survey that enables our measurements of the X-ray Luminosity Function (XLF).

The X-ray Telescope on the \textit{Neil Gehrels Swift Observatory} \citep[hereafter the \textit{Swift}-XRT;][]{Burrows2005} has performed both targeted observations of X-ray sources and searches for unknown X-ray counterparts to transient sources, often following the detection of Gamma-ray bursts by \textit{Swift}'s own Burst Alert Telescope (\textit{Swift}-BAT). \rredit{Thus, \textit{Swift}-XRT has obtained} imaging of nearly 4000~degrees$^2$ of sky throughout its observing life \citep[as of August 2018;][]{Evans2020}. The ExSeSS sample was formed using the second \textit{Swift}-XRT Point Source (2SXPS) catalogue of \citet{Evans2020} that identified all point sources in the full data set provided by \textit{Swift}-XRT observations between 2005 January 01 and 2018 August 01. The effect of contamination by Galactic sources and nearby, individually resolved sources in nearby galaxies was reduced by removing the areas of sky corresponding to the Galactic plane (Galactic latitudes $|b|<20^{\circ}$), the Large and Small Magellanic Clouds, M31 and M33, ensuring the sample is dominated by extragalactic sources. In addition, only fields identified as \textit{ultra-clean} (field flag$=$0) in 2SXPS are included in \rredit{the ExSeSS datasets} and only sources with a \textit{good} detection flag are included in the sample.

\rredit{In order to create a truly serendipitous sample and remove any sources that may bias the sample due to association with the target,}
all target objects were removed along with any associated X-ray detections using regions of the radius of the source (where an extended counterpart could be identified, e.g. a host galaxy) and adopting a minimum radius of 2~arcminute.
Medium (1-2~keV), hard (2-10~keV) and total (0.3-10~keV) energy band detections by \textit{Swift}-XRT were taken, and wherever there are multiple observations of sky only the stacked images are used in order to maximise the exposure time. This process to create the ExSeSS sample is detailed in \citet{DelaneyPrep}.

A key feature of ExSeSS is that \rredit{that the survey volume can be defined, enabling our goals to place constraints on the XLF. The} area of sky surveyed by \textit{Swift} to different exposure times is carefully tracked and well defined, enabling an accurate quantification of the ``area curve'', giving the area of sky covered \rredit{by ExSeSS }to different X-ray flux limits \rrredit{\citep[see figure 5 of][]{DelaneyPrep}}. 
\citet{DelaneyPrep} use simulations \citep[from ][]{Evans2020} to determine the area curves in the soft, hard and total energy bands. \rredit{They are calculated using the sky area coverage of the survey, excluding the areas within the specified radius of target sources and areas corresponding to the Galactic plane and nearby galaxies, thus matching the sample definition described above \citep[see][for details]{DelaneyPrep}.}
It is these area curves\rredit{, and the serendipitous nature of the survey, }that enable us to perform measurements of the X-ray Luminosity Function based on the ExSeSS sample. 

The resulting ExSeSS catalogue is comprised of 79\,152 X-ray sources and covers 2086.6~degrees$^2$ of sky. 
\begin{figure*}
    \centering
    \includegraphics[width=17cm]{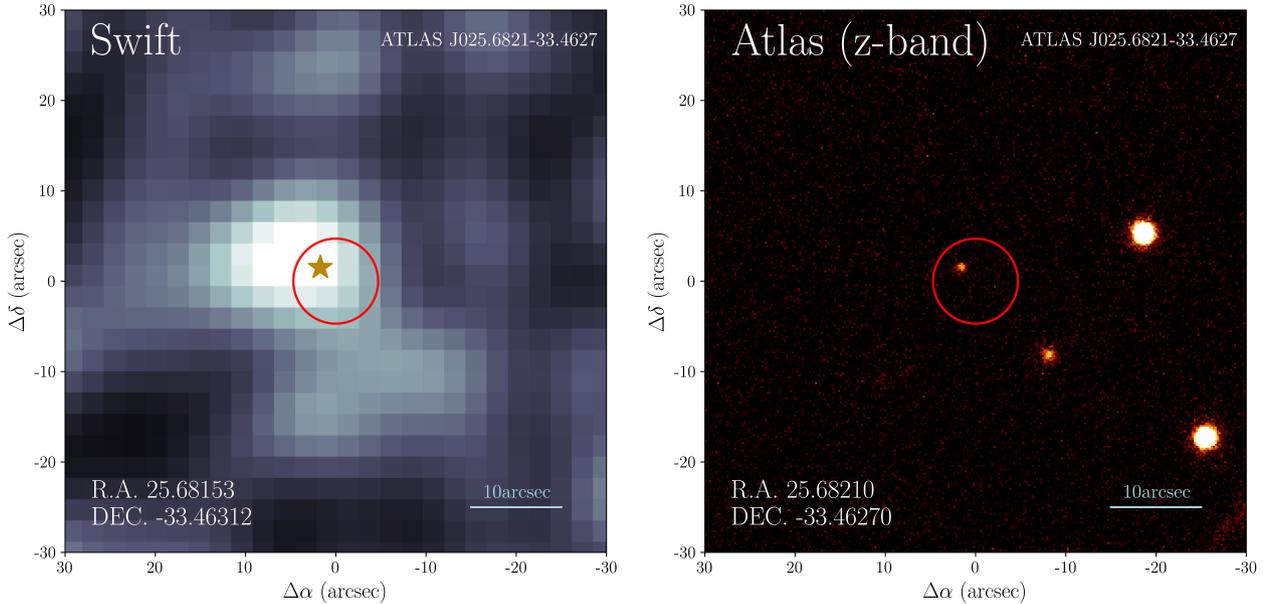}
    \caption{The high redshift AGN ATLAS J025.6821-33.4627, identified in ExSeSS. The stacked 0.3--10~keV X-ray image from Swift (left), smoothed by a Gaussian of $\sigma\sim2\,\mathrm{pixels}$ corresponding to the half-energy width of the PSF of Swift ($9\,\mathrm{arcsec}$), and the z-band Atlas image (right) are shown. The optical position of the source is shown on the Swift image by the gold star. 
    The radius of the solid red circles correspond to the 4.7~arcsec positional uncertainty of the source in Swift, centred on the observed soft-band position. \rredit{It is clear from these images that no additional z-band sources lie within the positional uncertainty of the X-ray source, indicating that} 
        \rredit{the association between ATLAS J025.6821-33.4627 and the \textit{Swift} X-ray detection is reliable}. The sky coordinates of the optical and X-ray sources are given in the corresponding image.}
    \label{fig:SourceImages}
\end{figure*}

\section{Identifying X-ray Luminous High-Redshift Sources}\label{sec:Sources}

In order to apply constraints to the high redshift X-ray Luminosity Function (XLF) with ExSeSS, we searched for all $z>5.7$ sources in ExSeSS. 

While a full statistical cross-matching between ExSeSS sources and multiwavelength surveys, to identify robust counterparts, will be presented in a future work, here we adopt a simple cross-matching to existing redshifts in the SIMBAD Database \citep{SimbadDatabase}. We use a maximum separation on the sky of 9~arcsec, given by the half-energy width of the point spread function of the \textit{Swift}-XRT \redit{(corresponding to the $2\sigma$ median positional uncertainty)} with $>90\%$ of all ExSeSS sources having a positional uncertainty of less than this value. \redit{18\,363 potential counterparts to the ExSeSS sources with pre-existing redshifts in SIMBAD were identified.} 

We note that the majority of ExSeSS sources do not have pre-existing counterparts or redshift measurements. Nevertheless, following this initial cross-matching we identified one high redshift ($z>6$) counterpart to the ExSeSS X-ray sources, that of 2SXPS J014243.7-332742, corresponding to the previously \rredit{optical/IR} detected quasar ATLAS J025.6821-33.4627. This source was then visually checked to ensure there are no other potential counterparts to the X-ray source. At near-infrared wavelengths ATLAS J025.6821-33.4627 appears as a point source, as can be seen in Figure~\ref{fig:SourceImages}, with a separation of 2.3~arcsec between the ExSeSS source and the counterpart and no other sources within the 4.7~arcsec uncertainty in the X-ray position. \redit{We assessed the probability of a spurious alignment using the number of sources in the \citet{Ross&Cross2020} catalogue over the total area covered by their sample to estimate the sky density of high-z AGN. This sky density is then multiplied by 
our search area, corresponding to a 9~arcsec radius around all 79 152 ExSeSS sources, to obtain the estimated probability of a spurious alignment. We find the probability of spurious alignment between an ExSeSS source and a known high-$z$ AGN to be only 0.03, indicating that the ExSeSS source and the counterpart identified are most likely the same source.}\footnote{\rrredit{\citet{Evans2020} estimate that $\lesssim0.3$\% of the ``good'' X-ray detections in 2SXPS, used to construct ExSeSS, are spurious X-ray detections. Thus the chance of a spurious X-ray detection \emph{and} spurious alignment with a known high-$z$ quasar is extremely low ($\lesssim0.009$\%).}}

\rrredit{ATLAS J025.6821-33.4627 is detected in a stacked dataset, comprised of 18 separate \textit{Swift}-XRT observations with a total exposure time of 189~ks \citep[see][for details]{Evans2020,DelaneyPrep}.}
The total (0.3--10~keV) band X-ray flux of this source was estimated from the measured total-band count rate, observed by the \textit{Swift}-XRT, using a conversion factor of $3.256\times10^{-11}\mathrm{erg~s^{-1}~cm^{-2}/(counts~s^{-1})}$ \footnote{We adopt a fixed flux conversion that assumes a Galactic absorption with a column density of $N_{\mathrm{H}}=2.50\times10^{20}\mathrm{erg~s^{-1}~cm^{-2}}$ \citep{HI4PIcollab2016} and a photon index of $\Gamma=1.9$ \citep[e.g.][]{Kalfountzou2014} calculated using \textit{WebPIMMS}, which matches the assumptions used to calculate the ExSeSS area curves.}\redit{, following the process in} \citet{DelaneyPrep}.
The rest-frame 2--10~keV X-ray luminosity was then determined from the observed 0.3--10~keV flux, assuming a power-law of photon index $\Gamma=1.9$. \rrredit{We note that the estimated X-ray luminosity of this source does not change significantly (compared to the quoted uncertainty based on the Poisson errors in the observed X-ray count rate) when assuming $\Gamma=$1.6 to 2.2 and thus this assumption is reasonable.} Given the high redshift of the source, the observed 0.3--10~keV band probes high rest-frame energies \rredit{($\sim$2.2--73.1~keV) and thus the observed flux would only be significantly suppressed by intrinsic column densities of $N_\mathrm{H}\gtrsim 10^{23}$~cm$^{-2}$.}
\rredit{Given that the source exhibits broad optical emission lines, it is unlikely to be heavily obscured at X-ray wavelengths and thus we have not applied any additional correction for intrinsic absorption when estimating the rest-frame 2--10~keV luminosity}. 
\rrredit{The source is only detected in the total 0.3--10~keV energy band, with 19.31 net counts, and thus we do not have sufficient constraints to make a direct estimate of photon index or absorption column.}
The 0.3--10~keV band observed flux and rest-frame 2--10~keV band luminosity, with the sky coordinates of this high redshift X-ray source and spectroscopic redshift of the counterpart, are given in Table~\ref{tab:SourceData}.

\begin{table} 
    \centering
    \caption{The \textit{Swift} X-ray position \rrredit{and observed X-ray properties of the high-redshift ExSeSS source, along with the \rrredit{ATLAS} z-band AB magnitude, spectroscopic redshift ($z_{spec}$), the rest-frame 2--10~keV luminosity calculated from the total-band flux, monochromatic luminosities at X-ray and optical wavelengths and the optical--to--X-ray slope, $\alpha_\mathrm{OX}$ (see Equation~\ref{eq:alphaox}).}
    }
    \begin{tabular}{c|c}
       Object & ATLAS J025.6821-33.4627 \\
        \hline
       RA (deg) & $25.68211^{+0.00081}_{-0.00073}$ \\
       Dec (deg) & $-33.46189^{+0.00057}_{-0.00056}$ \\ 
       Total 0.3--10~keV counts & $63$ \\
       Net 0.3--10~keV counts & $19.31$ \\
       0.3--10~keV count rate (cts~s$^{-1}$) & $1.4^{+0.6}_{-0.5} \times 10^{-4}$\\
       $f_{\mathrm{0.3-10keV}}$ ($\mathrm{erg\,s^{-1}\,cm^{-2}}$) & 
        $4.59^{+1.84}_{-1.69}\times10^{-15}$ \\ \hline
       z$_{ABmag}$ & $19.63\pm0.06$ \\
    $z_{spec}$ & $6.31\pm0.03$ \\
       $L_{\mathrm{X_{2-10keV}}}$ ($\mathrm{erg\,s^{-1}}$) &  $8.47^{+3.40}_{-3.12}\times10^{44}$ \\
       $L_{\mathrm{2keV}}$ (erg~s$^{-1}$~Hz$^{-1}$) & $2.01^{+0.80}_{-0.74}\times10^{27}$ \\
       $L_{\mathrm{2500\mbox{\AA}}}$ (erg~s$^{-1}$~Hz$^{-1}$) & $3.90^{+0.23}_{-0.21}\times10^{31}$ \\
       $\alpha_{\mathrm{OX}}$ & $-1.65^{+0.06}_{-0.09}$ \\ \hline
    \end{tabular}
    \label{tab:SourceData}
\end{table}

\begin{figure}
    \centering
    \includegraphics[width=8.75cm]{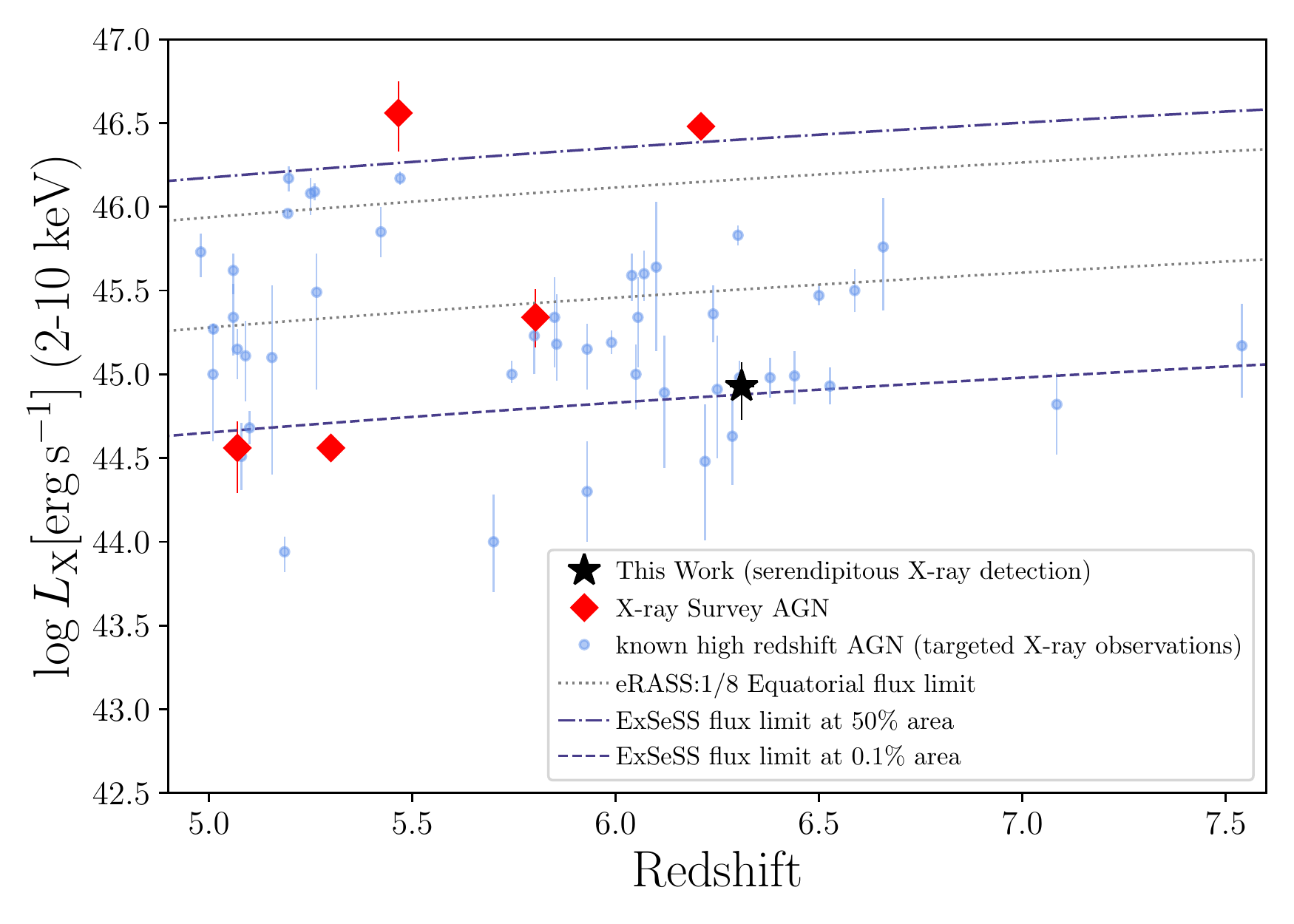}
    \caption{The X-ray luminosity of our $z>6$ ExSeSS source with respect to redshift (black star). Known high-redshift AGN with spectroscopic redshifts, \rredit{detected using targeted X-ray observations} \citep[see][and references therein]{Khorunzhev2021}, are shown by the light blue points. The few high-redshift sources \rredit{that are detected in the dedicated eROSITA and COSMOS X-ray survey fields} are shown for comparison (red diamonds). Our source, \rredit{which was serendipitously detected within ExSeSS}, is the highest redshift source identified by X-ray surveys with a well-defined area coverage from which an XLF can be determined. The sensitivity limits of \rrredit{ExSeSS, for 50\% and 0.1\% (the flux limit of ExSeSS) of the total area, are show by the dot-dashed and dashed purple lines respecitvely. For comparison, the eROSITA sensitivity limits are shown by the dotted grey lines, the upper line showing}  after 6 months (eRASS:1) \rrredit{and the lower showing the limit} after 4 years (eRASS:8), in \rredit{the equatorial region} \citep[][]{Predehl2021, Sunyaev2021}. Our ExSeSS $z>6$ quasar \rrredit{can be seen to} lie close to \rrredit{the sensitivity limit achieved by ExSeSS, below the eRASS limits,} and \rredit{is the highest redshift, serendipitously X-ray detected AGN to date.}}
        \label{fig:Distribution}
\end{figure}

ATLAS J025.6821-33.4627 was originally identified as a candidate high-redshift quasar by \citet{Carnall2015}, based on its combined WISE and VST ATLAS colours, indicative of a $z=5.7-6.4$ source. Follow-up spectroscopy was obtained, by \citet{Carnall2015}, using the Low Dispersion Survey Spectrograph 3 on the Magellan-II telescope from which a redshift of $z=6.31\pm0.03$ was calculated based on the broad Lyman-$\alpha$ line in the source's spectrum. 
The X-ray properties of ATLAS J025.6821-33.4627 from 2SXPS are included in the compilation of known high-redshift quasars by \citet{Vito2019} but we now identify this source as a serendipitous \rredit{X-ray }detection with ExSeSS: \rredit{we stress that ATLAS J025.6821-33.4627 was \emph{not} the target of the \textit{Swift} observation.}
Comparing to known X-ray detected high-redshift AGN \citep[see][and references there in]{Khorunzhev2021}, as shown in figure \ref{fig:Distribution}, ATLAS J025.6821-33.4627 can be seen to lie just above the limit of the deepest sensitivity expected with eROSITA (that obtained at the Polar regions after 4 years of the survey; eRASS:8). 
\rredit{Whilst X-ray detections of many $z>5$ AGN have been reported, most are the result of targeted X-ray observations; very few X-ray detections have been obtained \emph{serendipitously} or within a dedicated survey fields (see figure \ref{fig:Distribution}), either of which is required to have a well-defined survey area that can be translated into a survey volume, enabling measurements of space densities and the XLF.} With a rest-frame 2--10~keV X-ray luminosity of $L_\mathrm{X}=8.47^{+3.40}_{-3.13}\times10^{44}\mathrm{erg\,s^{-1}}$, in ExSeSS, ATLAS J025.6821-33.4627 is likely the highest redshift, spectroscopically confirmed, serendipitously \rredit{X-ray detected} quasar known to date.\footnote{\rrredit{Since submission of this paper, \citet{Wolf2022} have reported a low-significance X-ray detection of a quasar with a spectroscopic redshift of $z=6.56$ in the 140~deg$^2$ eFEDS field.}}

During the cross-matching of sources within ExSeSS a second bright X-ray source was matched to the previously identified Quasar 5C 2.183, with an spectroscopic redshift from the SDSS database of $z=6.16892\pm0.00060$ \citep[][]{Paris2017}. However, as this source was detected in u, g, r, i, and z-bands of SDSS as well as the G-band of Gaia, not possible in a $z\gtrsim5$ source, closer inspection of the SDSS spectrum of the source was performed. From this spectrum it can be seen that the redshift of the source is in fact $z=0.714$ as identified by other studies \citep[][]{Machalski1998, Paris2018, Chen2018} and thus this is not an additional high redshift AGN within ExSeSS.

With this contaminant removed, and no other high-redshift sources identified \redit{in ExSeSS}, we can give a tentative number of high redshift sources with X-ray luminosities high enough to be detectable by Swift-XRT at $z\gtrsim6$, \rredit{which we can compare to predictions based on model extrapolations and use to place constraints on the XLF (see \S~\ref{sec:XLFconstraints} below).
However, we first examine the X-ray and optical properties of ATLAS J025.6821-33.4627 in more detail.
} 

\subsection{Optical to X-ray Properties \rredit{of ATLAS J025.6821-33.4627}}\label{sec:OpticalAndBol,sub:alphaOX}

High-redshift AGN are often selected based on their optical and UV properties. 
Whilst ATLAS J025.6821-33.4627 is a highly luminous source at optical wavelengths, we have shown that it is identified serendipitously based on X-ray selection as part of ExSeSS. In order to investigate its nature and the relation between the emissions of its accretion disk and X-ray corona, we determine the optical-to-X-ray relation of the source, as in \citet{Vito2019} \rrredit{\citep[see also][]{Tananbaum1979}}, given by 
\begin{equation}
    \alpha_{\mathrm{OX}}=0.3838\log{\left(\frac{L_{\mathrm{2keV}}}{L_{\mathrm{2500\mbox{\AA}}}}\right)}
    \label{eq:alphaox}
\end{equation}
where the optical-to-X-ray slope, $\alpha_{\mathrm{OX}}$, is given by the ratio of the monochromatic X-ray luminosity of the source at a rest-frame energy of 2~keV, $L_{\mathrm{2~keV}}$, and the optical luminosity of the source at a rest-frame wavelength of  2500~\AA, $L_{\mathrm{2500\mbox{\AA}}}$. We determine the optical luminosity using the observed z-band luminosity and assuming a power-law continuum of $f_\nu\propto\nu^{\alpha_\nu}$, with $\alpha_\nu=-0.3$ \citep[see][and references therein]{Banados2016, Pons2020}, as detailed in appendix \ref{sec:Appendix}. The 2~keV luminosity is determined from the 2--10~keV band luminosity assuming a photon index of $\Gamma=1.9$ (see equation \ref{eq:MonoX-rayLuminosity} in the appendix). The values used are given in table \ref{tab:SourceData}.

\begin{figure}
    \centering
    \includegraphics[width=8.75cm]{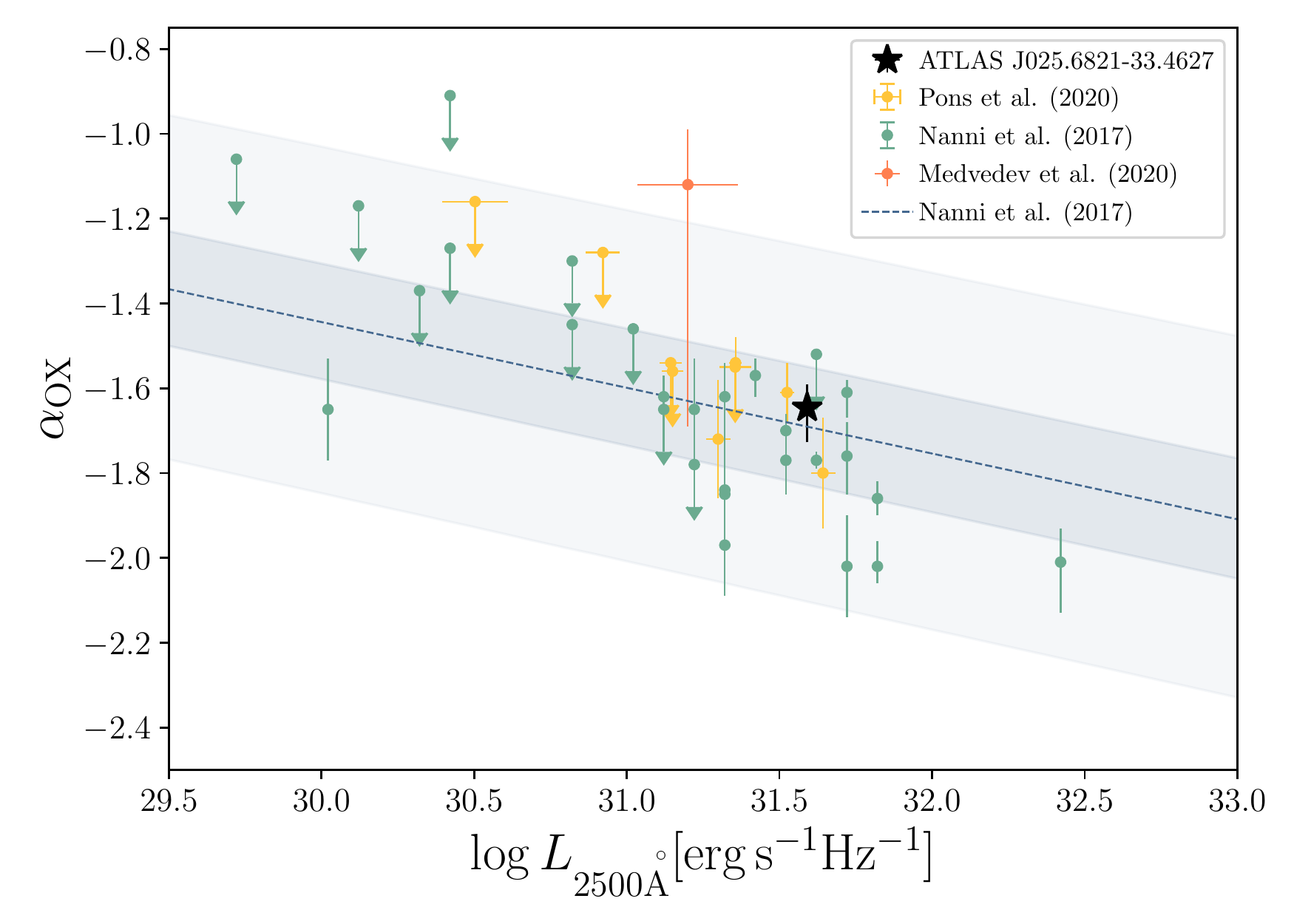}
    \caption{The optical-to-X-ray slope, $\alpha_\mathrm{OX}$, and 2500\AA\ monochromatic luminosity of the serendipitously detected ExSeSS source ATLAS J025.6821-33.4627 (black star). For comparison, measurements of $\alpha_{\mathrm{OX}}$ for X-ray-targetted samples of high-redshift AGN from \citet{Pons2020}, \rrredit{\citet{Medvedev2020}} and \citet{Nanni2017} are shown by the yellow, \rrredit{pink} and turquoise points, respectively. The standard relation for the optical-to-X-ray slope, as a function of the 2500\AA\ monochromatic luminosity, determined by \citet{Nanni2017}, is shown by the dashed line with shaded regions showing the $1\sigma$ (grey) and $3\sigma$ (light grey) scatter in the relation. Our high redshift ExSeSS source can be seen to \rredit{lie within the $1\sigma$ scatter} of the expected relation.}
    \label{fig:alpha_OX}
\end{figure}

We find that the optical-to-X-ray slope of ATLAS J025.6821-33.4627 is $-1.65^{+0.06}_{-0.09}$, which \rredit{lies within the $1\sigma$ scatter} of the $\alpha_{\mathrm{OX}}-L_{\mathrm{2500\mbox{\AA}}}$ relation of \citet{Nanni2017} (as shown in figure \ref{fig:alpha_OX}). Thus, although this is a relatively optically bright source, such an optical luminosity is \redit{consistent with that expected, given} its X-ray luminosity. This indicates that the \rredit{accretion mechanism} in this high redshift source is likely the same as seen in lower redshift AGN, as it follows the same optical-to-X-ray slope, and the source detection \rredit{in ExSeSS} is not a consequence of being relatively X-ray luminous. 
\redit{Despite being X-ray selected and optically bright, we find that \rredit{ATLAS J025.6821-33.4627} is typical of the AGN population observed at high redshift.} \rredit{Thus, the observed XLF constraints calculated including this source (see \S\ref{sec:XLFconstraints}) are expected to be indicative of the typical AGN population at high-z.}

\section{Observational constraints on the high-redshift XLF}\label{sec:XLFconstraints}

Using the \rredit{sample of serendipitously detected} high-redshift X-ray source\rredit{s} identified in ExSeSS, we calculate limits on the AGN space density and place constraints on the bright end of the high-redshift XLF. In \S\ref{sec:XLFconstraints,sub:numbers} we compare the number of sources observed with the number predicted by extrapolations of parametric XLF models. In \S\ref{sec:XLFconstraints,sub:XLF}, we present the constraints on the XLF that are obtained from the ExSeSS sample. \rredit{We then compare our X-ray based constraints at $z>6$ to prior measurements of the bolometric QLF, primarily based on optical samples at these redshifts, in \S\ref{sec:OpticalAndBol,sub:BolQLF}.}

As \redit{not all ExSeSS sources have counterparts} there may be additional unidentified high-redshift sources in ExSeSS, for which we do not have redshift information. \redit{Even if there is a significant population of unidentified AGN at high redshift, we are still able to place \emph{\redit{lower}} limits on the XLF based on the ExSeSS survey.} However, if we assume that any high-luminosity X-ray AGN would also be bright at rest-frame UV wavelengths \rredit{\citep[as predicted by the optical-to-X-ray relation see, e.g.][]{Nanni2017, Pons2020} and that they are unobscured} then we would expect such sources \rredit{to} have been identified in the numerous \rredit{UV/optical} searches for high-$z$ quasars \rrredit{\citep[that have now covered the majority of the extragalactic sky in both hemispheres, e.g.][]{Banados2016,Wang2019,Reed2019}} and as such would have entered our \redit{redshift} sample, if they exist \rredit{in ExSeSS}. \redit{\rredit{Thus,} we do \emph{not} expect there to be significant additional high-redshift AGN \rredit{within the ExSeSS X-ray selected sample} that we have not yet identified.}
\redit{Nevertheless, we note that the constraints obtained here are formally \emph{lower} limits only.} 
\subsection{Predicted numbers of {\ensuremath{z\gtrsim6}} AGN} \label{sec:XLFconstraints,sub:numbers}

\begin{figure}
    \includegraphics[width=8.75cm]{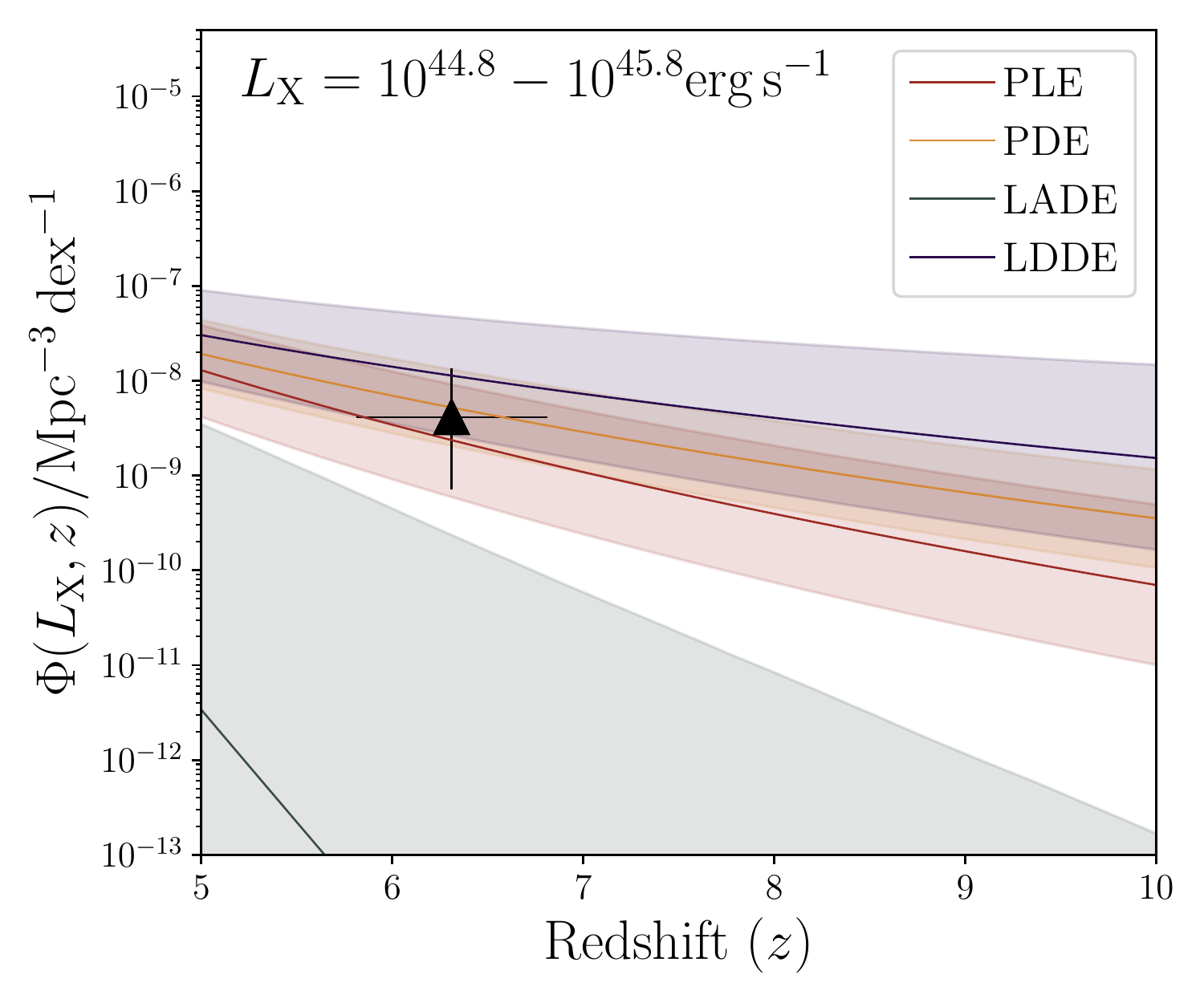}
    \caption{Measurement of the space density of AGN in the  $\log{L_{\mathrm{X}}}=44.8-45.8$ luminosity bin, \redit{based on the identification of a single high-redshift source in ExSeSS source and assuming no other high-redshift sources exist within ExSeSS (\rredit{formally a lower limit; black triangle})}. The space densities predicted by the \citet{Georgakakis2015} models as a function of redshift are also shown, with the shaded regions showing the $1\sigma$ uncertainties in the predictions, and are generally consistent the observational constraints.
        }
    \label{fig:SpaceDensities}
\end{figure}

The number of AGN X-ray sources at different luminosities and redshifts that are expected to be observed in an X-ray survey can be predicted using XLF models and the area curve of the survey. As the area curve of the ExSeSS survey has been calculated we can perform such predictions of the expected number of AGN in ExSeSS. 

Using the parametrised XLF models of Pure Luminosity Evolution (PLE), Pure Density Evolution (PDE), Luminosity And Density Evolution (LADE) and Luminosity Dependent Density Evolution (LDDE) from \citet{Georgakakis2015}, we calculate the expected number of sources in ExSeSS. The XLF models are extrapolated out to high redshifts, then the predicted number obtained using the integral 

\begin{equation}
    N_{model}=\int_{z_1}^{z_2}\int_{\log{L_{X_1}}}^{\log{L_{X_2}}} \phi(L_X,z)A(f(L_X,z))\frac{dV_\mathrm{co}}{dz}d\log{L_X}dz
    \label{eq:predictNumber}
\end{equation}

where the XLF, $\phi(L_X ,z)$, is the parametrised model \citep[PLE, PDE, LADE or LDDE from][]{Georgakakis2015}, $\frac{dV_\mathrm{co}}{dz}$ is the differential comoving volume and $A(f(L_X ,z))$ is the sky area  covered by ExSeSS to an observed flux, $f(L_X ,z)$, corresponding to a given 2--10~keV rest-frame luminosity, $L_X$, and redshift, $z$. \redit{We fix the redshift limits to $z_1=5.7$ and $z_2=6.4$, corresponding to the selection window of the \citet{Carnall2015} study \rredit{(based on i-band drop-out selection)}.} 
The $1\sigma$ uncertainties in these predicted numbers are obtained through Monte Carlo simulations using the model parameter uncertainties of \citet{Georgakakis2015}. \rredit{We choose not to perform a global correction to the area curve for the spectroscopic completeness of the ExSeSS sample as the completeness is likely to vary substantially with redshift in a poorly constrained manner.}

\redit{Our predicted numbers of sources in the ExSeSS survey, based on the  \citet{Georgakakis2015} models, are given in Table~\ref{tab:Predictions} and compared to our observed source numbers.}
\redit{We adopt 1~dex wide luminosity bins, with the minimum luminosity corresponding to the flux (for a source at $z>5.7$) where the area curve drops to $0.1\%$ of the total area of ExSeSS\rrredit{, assuming a spectral index of $\Gamma=1.9$}, in order to avoid the uncertainties inherent in the area curve at fainter fluxes}. This results in the lowest luminosity bin being $\log{L_\mathrm{X}}=44.8-45.8$, in which the source, 
ATLAS J025.6821-33.4627, identified in ExSeSS falls.
However, we find no sources in the 1~dex higher luminosity bin, $\log{L_\mathrm{X}}=45.8-46.8$, or at even higher luminosities, which we would expect to \redit{be optically brighter \rredit{\citep[given the optical-to-X-ray slope at high redshift;][]{Nanni2017}} and thus would have been easier to identify in prior \rredit{UV/optical }searches for high-$z$ quasars (assuming such high-luminosity sources are not obscured) \rredit{and hence fall into our high-$z$ population sample}}. Thus, we take an upper limit on the \redit{observed} number of sources, $N_{obs}$,  in this higher luminosity bin given by the upper $1\sigma$ equivalent Poisson limit for a sample size of $N=0$ from \citet{Gehrels1986}. 

\redit{In general, we find good agreement between our predicted and observed source numbers given in Table~\ref{tab:Predictions}, indicating that the XLF model \rredit{extrapolations} give reasonable predictions for the number of AGN at these luminosities and redshifts.
The PLE and PDE models generally predict $\sim1$ source in ExSeSS at $\log L_\mathrm{X}=44.8-45.8$ and $<1$ in the higher luminosity bin, consistent with our observed sample. The LDDE model predicts slightly higher numbers but remains consistent with the observed numbers.} 
However, the LADE model under-predicts the number of observed sources, albeit with a very large uncertainty.
Nonetheless, the $1\sigma$ upper limit based on the extrapolated uncertainty in the LADE model remains below the $1\sigma$ limit from our observed number; the LADE model, while not formally ruled out, is thus disfavoured based on our measurements. 

\begin{table}
    \centering
    \caption{The predicted number of sources, at $z=5.7-6.4$, based on the four XLF models from \citet{Georgakakis2015} in the two luminosity bins where we place constraints on the number of high-redshift sources using ExSeSS. The $1\sigma$ uncertainties on the model predictions are obtained through Monte Carlo error propagation (for the LADE model we give the $1\sigma$ upper limit only given the large range). The $N_{obs}/N_{model}$ binned XLF estimates obtained from the observed ExSeSS sources, $\phi(L_{\mathrm{X}}, z)$, and the observed number of AGN in each luminosity bin are also given, with $1\sigma$ limits based on the Poisson errors from \citet{Gehrels1986} (see \S\ref{sec:XLFconstraints,sub:XLF} for details).}
    \begin{tabular}{c|c|c}
        \hline
        Model & $\log{L_\mathrm{X}}=44.8-45.8$ & $\log{L_\mathrm{X}}=45.8-46.8$ \\ 
        & $\mathrm{erg~s^{-1}}$ & $\mathrm{erg~s^{-1}}$ \\ \hline
        PLE & $0.62^{+1.60}_{-0.48}$ & $0.30^{+1.07}_{-0.25}$ \\
        PDE & $1.05^{+1.48}_{-0.67}$ & $0.31^{+0.66}_{-0.22}$  \\ 
        LADE & $<0.051$ & $<0.015$ \\
        LDDE & $2.76^{+10.8}_{-2.25}$ & $1.80^{+16.8}_{-1.63}$ \\
        \hline
        $\phi(L_{\mathrm{X}}, z)$ & $2.98^{+6.86}_{-2.47}\times10^{-9}$ & $<2.22\times10^{-10}$ \\
        $1/V_{\mathrm{max}}$ & $3.27^{+7.52}_{-2.70}\times10^{-8}$ & No Data \\
        \hline
        $N_{obs}$ & $1^{+2.300}_{-0.827}$ & $<1.814$ \\
        \hline
    \end{tabular}
    \label{tab:Predictions}
\end{table}

\subsection{AGN space densities and measurements of the XLF}
\label{sec:XLFconstraints,sub:XLF}

\begin{figure*}
    \includegraphics[width=12cm]{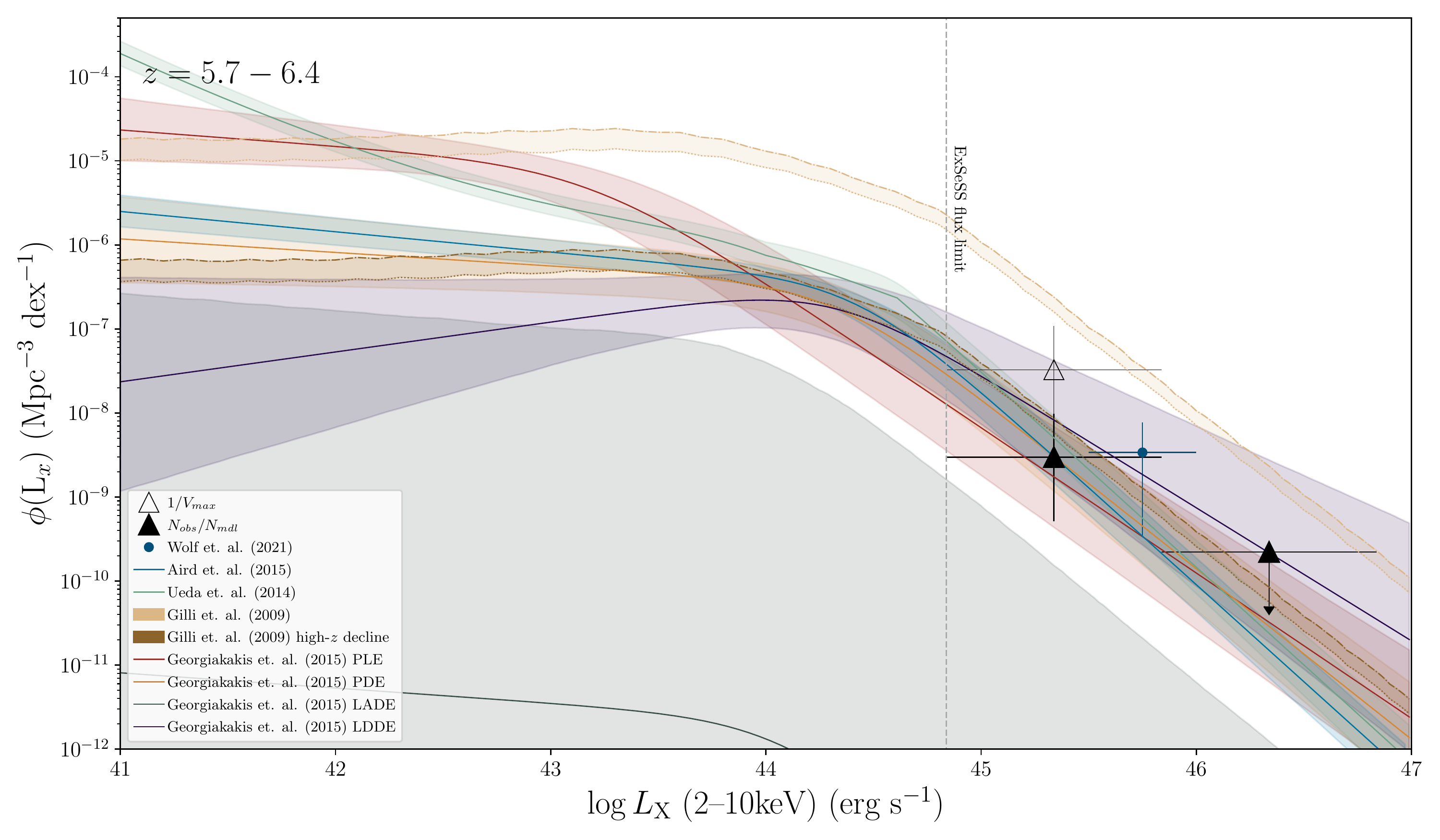}
    \caption{
    \rredit{Our measurements of the XLF at $z\sim6$ compared to extrapolations of the various different parameterised models from \citet{Georgakakis2015} and a number of other recent XLF models.}
    \rredit{Our} binned XLF measurements, given by the $N_{obs}/N_{model}$ method (solid \rredit{triangle}), are plotted at the centre of the luminosity bins in which they were calculated with the width of the luminosity bins shown by the horizontal error bars. Vertical error bars give the $1\sigma$ uncertainties on the XLF estimates. We also show a binned measurement using the ${1/V_{max}}$ method in the lower luminosity bin (hollow \rredit{triangle}) for comparison. The observed XLF found by \citet{Wolf2021} is shown by the blue circle, and can be seen to agree with our constraints to within $1\sigma$. \redit{We note that these binned XLF estimates are calculated assuming no other $z>5.7$ sources exist within ExSeSS \rredit{and thus are formally lower limits}.} 
    \redit{The detection limit of ExSeSS, given by the $0.1\%$ of the total observed area (as in \S\ref{sec:XLFconstraints,sub:numbers}), is shown by the vertical dotted line.}
            \rrredit{XLF models fitted by previous studies, extrapolated to the redshift range here, are plotted with $1\sigma$ uncertainties shown by a shaded region. The PLE (red), PDE (yellow), LDDE (purple) models of \citet{Georgakakis2015}, \citet{Ueda2014} LDDE model (green), \citet{Aird2015} Flexible Double Power-Law (FDPL; blue) model, and \citet{Gilli2007} LDDE with high L$_X$ decline model (dark brown) are consistent with the binned measurements. In contrast, the \citet{Georgakakis2015} LADE model and \citet{Gilli2007} model without the high L$_X$ decline (light brown) predict lower and higher space densities, respectfully, than has been observed with ExSeSS and are thus disfavoured.}
    Both \citet{Gilli2007} models are shown with (dot-dashed) and without (dotted) Compton-thick AGN. 
    }
    \label{fig:binnedXLF}
\end{figure*}

Extrapolating parametrised models of the AGN XLF, such as those of \citet{Georgakakis2015, Aird2015, Ueda2014}, out to high redshift, provides insights into the AGN population in the very early Universe. However, due to the lack of observational data at redshifts of $z>6$\rredit{, these models are not constrained at high redshifts and thus the predictions are based on extrapolating a given paremetric form (determined by the lower-$z$ data)}. \rredit{Even with just one high-redshift AGN found in ExSeSS, we can place new \emph{observational} constraints} on the space density of the high-redshift AGN population and compare with these models. 

We use the $N_{obs}/N_{model}$ method of \citet{Miyaji2001} to convert the observed number of sources in the luminosity bins of $\log{L_\mathrm{X}}=44.8-45.8$ 
into a measurement of the AGN space density. 
\redit{Our estimate of the space density, $\Phi_{est}(L_\mathrm{X},z_i)$, is}
calculated by scaling the predicted space density based on a given model of the XLF, $\Phi_{model}\left(L_{\mathrm{X}_i},z_i\right)$, by the ratio of the observed number of sources to the predicted number of sources, $N_{obs}/N_{model}$. Thus,
\begin{equation}
    \Phi_{est}\left(L_{\mathrm{X}_i}, z_i\right)=\Phi_{model}\left(L_{\mathrm{X}_i}, z_i\right)\frac{N_{obs}}{N_{model}}
     \label{eq:spaceDensity}
\end{equation}
where the predicted number, $N_{model}$ is estimated using the LDDE model \citep[from][]{Georgakakis2015}, \rredit{using eq.\ref{eq:predictNumber},} for a redshift range of $z=5.7-6.4$ and in the luminosity bins given in Table \ref{tab:Predictions}, and the model space density, $\Phi_{model}\left(L_{\mathrm{X}_i}, z_i\right)$, is given by the integral of the LDDE model over the luminosity bin.  
The $1\sigma$ uncertainties on $\Phi_{est}\left(L_{\mathrm{X}_i}, z_i\right)$ are based on the Poisson uncertainties in the observed source number, as given by \citet{Gehrels1986}.

Our measurement of the space density of AGN with $\log L_\mathrm{X}=44.8-45.8$ is shown in figure \ref{fig:SpaceDensities} and compared to the space densities predicted by the four \citet{Georgakakis2015} XLF models. 
The measured space density is generally in good agreement with the extrapolated XLF models. The PLE and PDE models and the lower limit on the LDDE model agree
with the observed space density,  within the $1\sigma$ uncertainties, whilst the upper bound of the LADE model only agrees with the observed space density within the $3\sigma$ uncertainties \rredit{(not shown)}. Thus, our observed space density can be seen to show the expected decline in space density of AGN\rredit{, based on lower redshift data,} towards higher redshift, as seen in in the PLE, PDE and LDDE models, but this decline is may not be as rapid as predicted by the some models, such as LADE. 

Following a similar process, we also determine binned measurements of the XLF, given by
\begin{equation}
    \phi_{est}\left(L_{\mathrm{X}_i}, z_i\right)=\frac{d\Phi_{est}\left(L_{\mathrm{X}_i},z_i\right)}{d\log{L_{\mathrm{X}}}}=\frac{d\Phi_{model}\left(L_{\mathrm{X}_i},z_i\right)}{d\log{L_{\mathrm{X}}}} \frac{N_{obs}}{N_{model}}
\end{equation}
where the predicted number, $N_{model}$ is given by the LDDE model as in equation \ref{eq:spaceDensity}, $N_{obs}$ is the observed number of AGN and $\phi_{model}\left(L_{\mathrm{X}_i}, z_i\right)$ is taken to be the value of the LDDE model at the centre of the adopted redshift and luminosity bins. 
\redit{
Figure~\ref{fig:binnedXLF} presents these binned measurements in both the $\log L_\mathrm{X}=44.8-45.8$ bin where our single high-redshift detection lies. The $1\sigma$ uncertainties are based on the Poisson uncertainties in the observed source numbers, as given by \citet{Gehrels1986}.}
\rrredit{We also show an upper limit in the higher $\log L_\mathrm{X}=45.8-46.8$ bin where no sources are found, based on the $1\sigma$ upper limit (given zero detected sources) from \citet{Gehrels1986}. This constraint relies on the assumption that any sources with such high X-ray luminosities at these redshifts would also be optically bright and thus would have been identified in existing optical/IR quasar searches, but it should not be considered a stringent upper limit given the potential for obscured or optically weak sources within ExSeSS that remain unidentified.}

For comparison we also provide an XLF measurement in the lower luminosity bin based on the more commonly used $1/V_{max}$ method \citep[][]{Schmidt1968}. 
The $1/V_{max}$ method does not assume an underlying parametric model but is more biased than the $N_{obs}/N_{model}$ method as it does not account for a change in the XLF across a broad luminosity range or with redshift and is thus more strongly affected by source luminosity, in particular for low sample sizes. However, we find that our $N_{obs}/N_{model}$ estimate shows negligible change depending on the assumed XLF model and agrees well (within $1\sigma$) with the $1/V_{max}$ estimate.

As can be expected from figure \ref{fig:SpaceDensities}, our binned XLF measurements are found to be consistent with the fiducial values of the XLF models (see figure \ref{fig:binnedXLF}). The PLE, PDE and LDDE models from \citet{Georgakakis2015} agree with the observations within the $1\sigma$ Poisson uncertainty, whilst the LADE model in particular falls much lower than the binned XLF estimate (see \S\ref{fig:binnedXLF}). In addition to the \citet{Georgakakis2015} models, we also compare with the model XLFs of \citet{Aird2015} and \citet{Ueda2014}, which agree to within $1\sigma$ of the $\log{L_\mathrm{X}}=44.8-45.8$ binned XLF estimates, and \cite{Gilli2007} which lies more than $3\sigma$ above our binned XLF estimates, as shown in figure \ref{fig:binnedXLF}. The gradient of the bright-end of these XLF models is consistent with the gradient indicated by our binned XLF estimates, with a value of $\gamma\gtrsim0.367$, consistent with the relatively steep bright-end slope of the XLF at $z\gtrsim6$.

In Figure~\ref{fig:binnedXLF}, we also compare with the result obtained from eFEDs data by \citet{Wolf2021}. We can see that our measurements are consistent to within $1\sigma$ of the XLF value based on the single $z=5.81$ source found in the $\sim$140~deg$^2$ eFEDS survey.

As noted above, there may be other $z\gtrsim6$ AGN within ExSeSS that we have not identified as they do not currently have redshift estimates. \redit{There may also be a significant population of obscured AGN, which would result in a higher space density than observed here.} \redit{However, given the agreement between our observed XLF and the model \rredit{extrapolations}, additional as yet unidentified AGN are not expected. Regardless}, from the results presented here it can be seen that the ExSeSS \rredit{sample} provides important constraints on the bright-end of the high-redshift XLF. \redit{Further, significantly deeper observations would be needed to probe the faint-end of the \rredit{high-$z$} XLF\rredit{, where the model extrapolations can be seen to diverge,} \redit{and constrain the form of the XLF at high $z$.}}
\subsection{The Bolometric QLF}\label{sec:OpticalAndBol,sub:BolQLF}

\begin{table*}
    \centering
    \caption{Estimates of the bolometric luminosity for our source, calculated from the X-ray and 1450~\AA\ luminosities. The optical luminosity at wavelength of 1450~\AA\ is derived from the z-band luminosities, assuming a constant power-law relation of slope $\alpha_{\nu}=-0.3$, the bolometric luminosities are then calculated, from the 2-10~keV band and 1450~\AA\ wavelength luminosities, using the conversion method of \citet{Shen2020}.}
    \begin{tabular}{c|c|c|c}
        \hline
        Object &  $L_{\mathrm{bol\,from\,2-10\,keV}}$ & $L_{\mathrm{1450\mbox{\,\AA}}}$ &  $L_{\mathrm{bol\,from\,1450\mbox{\,\AA}}}$ \\ 
        & $\mathrm{erg\,s^{-1}}$ & $\mathrm{erg\,s^{-1}\,\mbox{\AA}^{-1}}$ & $\mathrm{erg\,s^{-1}}$ \\
        \hline
        ATLAS J025.6821-33.4627 & $1.54^{+0.91}_{-0.72}\times10^{47}$ & $2.52^{+0.15}_{-0.14}\times10^{45}$ & $2.29^{+0.13}_{-0.12}\times10^{46}$ \\ \hline
    \end{tabular}
    \label{tab:bolometricLuminosities}
\end{table*}

More high redshift AGN have been identified through rest-frame UV selection than X-ray selection, due to the limited survey areas covered with current X-ray telescopes compared to the efficiency and availability of large-scale optical/near-infrared imaging campaigns. In order to determine how well our observed XLF estimate compares to the luminosity functions given by \rredit{the more biased} Optical/UV selected AGN \rredit{samples}, we compare our X-ray binned XLF estimates to \rredit{models of} the bolometric QLF.

\begin{figure}
    \centering
    \includegraphics[width=8.75cm]{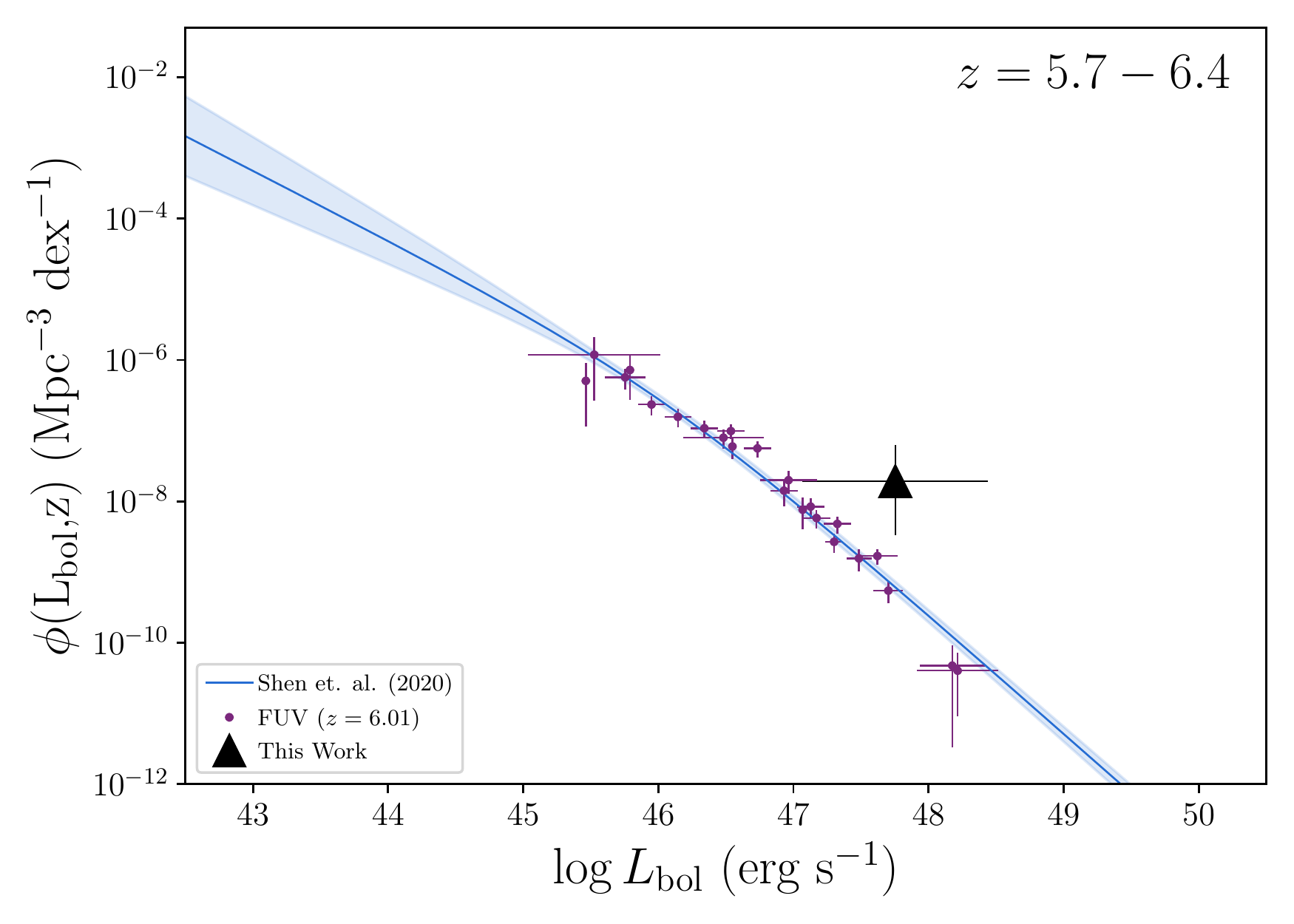}
    \caption{Binned estimate of the bolometric QLF (black \rredit{triangle}), converted from the binned XLF estimate found in \S\ref{sec:XLFconstraints,sub:XLF} based on the single source detection at $\log{L_\mathrm{X}}=44.8-45.8$\redit{ and assuming that there are no other high redshift sources in ExSeSS}. The estimate is calculated using the bolometric conversion factor of \citet{Shen2020}. \rredit{While our estimate lies above the \citet{Shen2020} QLF model (blue), it is consistent at the 2$\sigma$ level (1$\sigma$ error bar is shown).} Violet points show measurements at $z\sim6$ based on rest-frame UV data, converted to bolometric values, which were used to constrain the QLF model \citep[see][and references therein]{Shen2020}.
        }
    \label{fig:bolometricQLF}
\end{figure}

We convert the X-ray luminosity bins and binned XLF estimates into bolometric terms using a bolometric correction factor, adopting the luminosity-dependent corrections determined by \citet{Shen2020}, where $L_\mathrm{bol} = k_\mathrm{bol}(L_\mathrm{bol})\times L_\mathrm{X-ray}$ and $k_\mathrm{bol}(L_\mathrm{bol})$ is the luminosity-dependent X-ray-to-bolometric correction factor.
The binned XLF estimate \rredit{from the single source detection at $\log{L_\mathrm{X}}=44.8-45.8$} \rredit{is} converted to a bolometric QLF value and shown in figure \ref{fig:bolometricQLF} compared to the \citet{Shen2020} bolometric QLF model, which was based on a fit to a combination of bolometrically corrected UV/optical, IR and X-ray luminosity functions spanning a wide range of redshifts. At $z\sim6$, the \citet{Shen2020} model is primarily constrained by rest-frame UV selected data; the original measurements, compiled from the literature, are shown by the purple points in figure~\ref{fig:bolometricQLF} \citep[see][and references therein]{Shen2020}.

From figure \ref{fig:bolometricQLF} it can be seen that
our X-ray based estimates of the bolometric QLF are consistent to within $2\sigma$ of the \citet{Shen2020} model and the prior UV/optical measurements. \rredit{Similarly, the number of AGN within ExSeSS predicted by the \citet{Shen2020} QLF model (converted to an XLF and using equation~\ref{eq:predictNumber}) is $0.049^{+0.095}_{-0.0086}$, which is consistent to within $2\sigma$ with our observed number of sources in this bin, $N_{obs}=1^{+2.300}_{-0.827}$.}

\rredit{While our X-ray--based estimate of the bolometric QLF is consistent with the \citet{Shen2020} QLF model, we have adopted large X-ray--to--bolometric corrections at these luminosities \citep[as suggested by][]{Shen2020}.}
\rrredit{As discussed in \S\ref{sec:OpticalAndBol,sub:alphaOX}, our serendipitously detected AGN lies on the fiducial $\alpha_\mathrm{OX}$ -- $L_{2500\mbox{\AA}}$ relation (see Figure~\ref{fig:alpha_OX}) and thus appears typical in terms of its X-ray to optical properties for sources of such luminosity. However, at such high optical luminosities, the typical AGN is relatively X-ray weak due to the steep dependence of $\alpha_\mathrm{OX}$ on optical luminosity (i.e. a large fraction of their power is emitted at optical wavelengths), which indicates that a large X-ray--to--bolometric correction factor is required.}

\rrredit{At such high redshifts, inverse Compton scattering of cosmic microwave background photons from a jet, if present, can boost the X-ray emission from AGN \citep[see e.g.][]{Medvedev2020}, which would then lead to an over-estimate of the bolometric luminosity from the observed X-ray luminosity. However, our source has not been detected at radio wavelengths \citep[including with recent, deeper coverage from the Rapid ASKAP continuum survey:][]{Hale2021} and thus there is no evidence that a jet is present or that the X-ray emission is being boosted by non-coronal processes.}

\rredit{To further check whether such large  X-ray--to--bolometric corrections are warranted in this  luminosity--redshift regime, we also directly compare estimates of the bolometric luminosity of our source, ATLAS J025.6821-33.4627, based on the observed X-ray and rest-frame UV luminosities.}
We estimate the 1450\AA\ luminosity (based on the z-band magnitude of the source and assuming a power-law UV spectrum, as in \S\ref{sec:OpticalAndBol,sub:alphaOX}), which we use to estimate the bolometric luminosity of our source and compare with the bolometric luminosity derived from the observed X-ray luminosity \citep[adopting the relevant luminosity-dependent bolometric corrections from][in both cases]{Shen2020}. 
The various luminosity estimates are provided in table \ref{tab:bolometricLuminosities}.
The bolometric luminosity inferred from the rest-frame UV light is $L_\mathrm{1450\mbox{\AA}\; to \; bol} = 2.29^{+0.13}_{-0.12}\times10^{46}$~erg\,s$^{-1}$
which is consistent \redit{to within $2\sigma$ of} the bolometric luminosity inferred from the X-ray, $L_\mathrm{X-ray\; to \;bol} = 1.54^{+0.91}_{-0.72} \times 10^{47}$~erg\,s$^{-1}$ obtained using the large X-ray bolometric correction factor of $k_\mathrm{bol}=182.4$. 
\rredit{Thus, the large X-ray--to--bolometric corrections that we have adopted to convert our XLF estimates and compare with the bolometric QLF are warranted, given the properties of the single high-$z$ source in our ExSeSS sample.}

\section{Conclusions}\label{sec:Conclusions}

Luminosity functions provide a tracer of the AGN population across cosmic time. However, at high redshift these luminosity functions remain poorly constrained. \rredit{In this paper, we present observational constraints on the X-ray Luminosity Function (XLF) from the new Extragalactic Serendipitous Swift Survey \citep[ExSeSS;][]{DelaneyPrep}.}

\rredit{We} identified one X-ray selected AGN at $z>6$ within the carefully constructed sample of serendipitous X-ray sources from Swift-XRT observations that form ExSeSS \citep{DelaneyPrep}. 
The serendipitous X-ray source, 2SXPS J014243.7-332742, is matched with the optically bright $z=6.31\pm0.03$ quasar ATLAS J025.6821-33.4627, making it the highest redshift serendipitously \rredit{X-ray detected} quasar to date. With this detection, we are able to determine the space density of AGN and place constraints on the X-ray luminosity function at $z\sim6$, \redit{under the assumption there are no other high redshift sources within ExSeSS}.

Our conclusions are as follows:

\begin{itemize}
    \item 
The space density of AGN given by the high-redshift ExSeSS source shows the steep decline of AGN with increasing redshift. This observed decline is consistent with the expected exponential decline in the space density of luminous AGN with increasing redshift.
\redit{However, we note that any additional sources at these luminosities and redshifts that remain unidentified within ExSeSS, such as a significant population of obscured AGN, would result in a higher space density.}

\item
We place constraints on the shape of the $z\sim6$ XLF, \redit{assuming no other $z>5.7$ sources exist within ExSeSS, using} our single detection and an upper limit \rredit{on the number of sources} at higher luminosities. Combined, these constraints are consistent with a relatively steep bright-end slope of the XLF, with $\gamma\gtrsim0.367$, as found at lower redshifts. Our binned XLF estimates are broadly consistent with the extrapolation of parametric XLF models, based on fits to lower redshift data. However the constraints we have obtained here rule out the most extreme declines in the bright-end of the XLF indicated by some model extrapolations \citep[e.g. the LADE model of][]{Georgakakis2015}. \rredit{Furthermore, as there may be other high-$z$ sources in ExSeSS these constraints are formally lower limits, strengthening this conclusion.}

\item
Our XLF constraints are consistent (within $\sim 1\sigma$) with the prior measurements by \citet{Wolf2021}, which were based on the identification of a single $z=5.81$ X-ray selected AGN in the 140~deg$^2$ eFEDS field observed by eROSITA.

\item
Applying a bolometric correction to \rredit{our XLF measurement}, we find a good agreement with the parametric QLF model of \citet{Shen2020}. Our estimate of the bolometric QLF is consistent\rredit{, to within $2\sigma$,} with the QLF derived from rest-frame UV selected quasars at $z\sim6$.

\item
We find that the optical-to-X-ray slope, $\alpha_{\mathrm{OX}}$, of our serendipitously detected source is \redit{consistent to within $1\sigma$} of the expected $\alpha_{\mathrm{OX}}$-L$_{\mathrm{2500\mbox{\AA}}}$ relation, indicating  the accretion process in this high redshift source operates in a similar manner to AGN at lower redshift. \redit{Thus, despite being X-ray selected and optically bright, this source is typical of the population at this redshift. In addition,} as this source is optically bright, a high correction factor is required to converting its X-ray luminosity to a bolometric estimate.

\end{itemize}

The measurements presented in this paper provide important constraints on the extent of supermassive black hole growth within the early Universe. Our estimates, based on the 2086.6~degrees$^2$ covered by ExSeSS, indicate that the ongoing all-sky surveys being performed by eROSITA will identify a few tens to a few hundred high-luminosity AGN at $z>6$, once sufficient depth is achieved (i.e. by eRASS:8 all-sky depth), enabling further investigation of the growth of SMBHs within the early Universe. 
However, much deeper surveys will be required in order to constrain the faint end of the XLF \rredit{at high-$z$}, which will become possible within the next decade with new telescopes such as the \textit{Athena} X-ray Observatory.

\section*{Acknowledgements}

We thank the anonymous referees for comments that improved this paper. 
CLBH and JD acknowledge support from STFC PhD studentships.
JA acknowledges support from a UKRI Future Leaders Fellowship (grant code: MR/T020989/1). This research has made used of Swift-XRT data from the Neil Gehrels Swift Observatory and benefited greatly from the work by the Swift Team and the 2SXPS catalogue of \citet{Evans2020}. This research has made use of the SIMBAD database, operated at CDS, Strasbourg, France. We have benefited from the publicly available programming language {\sc Python}, including {\sc NumPy} \& {\sc SciPy} \citep[][]{NumPy, SciPy2020}, {\sc Matplotlib} \citep[][]{Matplotlib}, {\sc Astropy} \citep[][]{Astropy2013} and the {\sc Topcat} analysis program \citep{TopCat2013}. This research has made use of the online count rate simulator {\sc WebPIMMs} \citep[][]{PIMMS}.

\section*{Data Availability}

The ExSeSS catalogue and area curve used in this work is provided by \citet{DelaneyPrep} which is based on the 2SXPS catalogue \citep{Evans2020} available at \url{https://www.swift.ac.uk/2SXPS/} and provides the full X-ray data. ATLAS observations are available through the VST ATLAS public survey \url{http://osa.roe.ac.uk/#dboverview_div}.

\input{ms.bbl}

\appendix

\section{Derived Luminosity Calculations} \label{sec:Appendix}

Assuming the flux of the AGN follows a power law relation of $f_\nu\propto \nu^{\alpha_\nu}$ \citep[see e.g.][]{Pons2020, Banados2016, Selsing2016}, with $\alpha_\nu=-0.3$, the flux in the z-band can be converted to a different wavelength flux using

\begin{equation} \label{eq:FluxConversion}
    \frac{f_{\nu}}{f_{\mathrm{z-band}}}= \left(\frac{\nu_{\mathrm{obs}}}{\nu_{\mathrm{z-band}}}\right)^{\alpha_{\nu}} = \left(\frac{\lambda_{\mathrm{z-band}}}{\lambda_{\mathrm{rest}}(1+z)}\right)^{\alpha_{\nu}}
\end{equation}

where $f_{\nu}$ is the monochromatic flux at a rest frame frequency $\nu$ and $f_{\mathrm{z-band}}$ is the flux in the z-band (in units of $\mathrm{erg\,s^{-1}\,cm^{-2}\,Hz^{-1}}$), given by the observed z-band apparent magnitude. The ratio of these fluxes is given by the frequency of the z-band, $\nu_{\mathrm{z-band}}$, and the observed frequency at which to determine the monochromatic flux, $\nu_{\mathrm{obs}}$, or by the wavelength of the z-band, $\lambda_{\mathrm{z-band}}$, and the rest frame wavelength at which to determine the monochromatic flux, $\lambda_{\mathrm{rest}}$. For ATLAS the z-band central wavelength, used in this calculation, is 8780.0~\AA.

From the monochromatic flux, found using equation \ref{eq:FluxConversion}, the monochromatic luminosity (in units of $\mathrm{erg\,s^{-1}\,cm^{-2}\,\mbox{\AA}^{-1}}$) is given by the equation

\begin{equation} \label{eq:MonoOpticalLuminosity}
    L_{\lambda_{\mathrm{rest}}}=\frac{c}{\lambda_{\mathrm{rest}}^2} \frac{4\pi D_L^2}{\left(1+z\right)} \left(\frac{\lambda_{\mathrm{z-band}}}{\lambda_{\mathrm{rest}}\left(1+z\right)}\right)^{\alpha_{\nu}} f_{\mathrm{z-band}}
\end{equation}

where the monochromatic luminosity, $L_{\lambda_{\mathrm{rest}}}$, at a rest frame wavelength of $\lambda_{\mathrm{rest}}$ is related to the monochromatic flux observed in the z-band, $f_{\mathrm{z-band}}$, by the luminosity distance $D_L$, the rest-frame wavelength, the central wavelength of the z-band $\lambda_{\mathrm{z-band}}$ and the power $\alpha_\nu$. $z$ is the redshift of the source and $c$ is the speed of light.

The monochromatic luminosity at an energy of 2~keV, $L_{\mathrm{2\,keV}}$, is given by the equation

\begin{equation} \label{eq:MonoX-rayLuminosity}
    L_{E}=N(E)E=\frac{(2-\Gamma)L_{\mathrm{2-10~keV}}}{(10.0~\mathrm{keV}~^{2-\Gamma}-2.0~\mathrm{keV}~^{2-\Gamma})}E^{1-\Gamma}
\end{equation}

where the X-ray spectrum is assumed to be given by a power-law $N(E)\propto E^{-\Gamma}$, with a photon index of $\Gamma=1.9$, and the measured hard-band luminosity is given by the total (0.3-10~keV) band flux. The resulting monochromatic luminosity can then be converted from units of $\mathrm{erg.s^{-1}.keV^{-1}}$ to $\mathrm{erg\,s^{-1}\,Hz^{-1}}$ by multiplying the luminosity by a factor of h. For the calculation of $\alpha_{\mathrm{OX}}$, we calculate this monochromatic luminosity at an energy $E=\mathrm{2~keV}$

\bsp	\label{lastpage}

\begin{thebibliography}{}
\makeatletter
\relax
\def\mn@urlcharsother{\let\do\@makeother \do\$\do\&\do\#\do\^\do\_\do\%\do\~}
\def\mn@doi{\begingroup\mn@urlcharsother \@ifnextchar [ {\mn@doi@}
  {\mn@doi@[]}}
\def\mn@doi@[#1]#2{\def\@tempa{#1}\ifx\@tempa\@empty \href
  {http://dx.doi.org/#2} {doi:#2}\else \href {http://dx.doi.org/#2} {#1}\fi
  \endgroup}
\def\mn@eprint#1#2{\mn@eprint@#1:#2::\@nil}
\def\mn@eprint@arXiv#1{\href {http://arxiv.org/abs/#1} {{\tt arXiv:#1}}}
\def\mn@eprint@dblp#1{\href {http://dblp.uni-trier.de/rec/bibtex/#1.xml}
  {dblp:#1}}
\def\mn@eprint@#1:#2:#3:#4\@nil{\def\@tempa {#1}\def\@tempb {#2}\def\@tempc
  {#3}\ifx \@tempc \@empty \let \@tempc \@tempb \let \@tempb \@tempa \fi \ifx
  \@tempb \@empty \def\@tempb {arXiv}\fi \@ifundefined
  {mn@eprint@\@tempb}{\@tempb:\@tempc}{\expandafter \expandafter \csname
  mn@eprint@\@tempb\endcsname \expandafter{\@tempc}}}

\bibitem[\protect\citeauthoryear{{Aird}, {Coil}, {Georgakakis}, {Nandra},
  {Barro}  \& {P{\'e}rez-Gonz{\'a}lez}}{{Aird} et~al.}{2015}]{Aird2015}
{Aird} J.,  {Coil} A.~L.,  {Georgakakis} A.,  {Nandra} K.,  {Barro} G.,
  {P{\'e}rez-Gonz{\'a}lez} P.~G.,  2015, \mn@doi [\mnras]
  {10.1093/mnras/stv1062}, \href
  {https://ui.adsabs.harvard.edu/abs/2015MNRAS.451.1892A} {451, 1892}

\bibitem[\protect\citeauthoryear{{Astropy Collaboration} et~al.,}{{Astropy
  Collaboration} et~al.}{2013}]{Astropy2013}
{Astropy Collaboration} et~al., 2013, \mn@doi [\aap]
  {10.1051/0004-6361/201322068}, \href
  {https://ui.adsabs.harvard.edu/abs/2013A&A...558A..33A} {558, A33}

\bibitem[\protect\citeauthoryear{{Ba{\~n}ados} et~al.,}{{Ba{\~n}ados}
  et~al.}{2016}]{Banados2016}
{Ba{\~n}ados} E.,  et~al., 2016, \mn@doi [\apjs] {10.3847/0067-0049/227/1/11},
  \href {https://ui.adsabs.harvard.edu/abs/2016ApJS..227...11B} {227, 11}

\bibitem[\protect\citeauthoryear{{Ba{\~n}ados} et~al.,}{{Ba{\~n}ados}
  et~al.}{2018}]{Banados2018}
{Ba{\~n}ados} E.,  et~al., 2018, \mn@doi [\nat] {10.1038/nature25180}, \href
  {https://ui.adsabs.harvard.edu/abs/2018Natur.553..473B} {553, 473}

\bibitem[\protect\citeauthoryear{{Boyle}, {Shanks}, {Croom}, {Smith}, {Miller},
  {Loaring}  \& {Heymans}}{{Boyle} et~al.}{2000}]{Boyle2000}
{Boyle} B.~J.,  {Shanks} T.,  {Croom} S.~M.,  {Smith} R.~J.,  {Miller} L.,
  {Loaring} N.,   {Heymans} C.,  2000, \mn@doi [\mnras]
  {10.1046/j.1365-8711.2000.03730.x}, \href
  {https://ui.adsabs.harvard.edu/abs/2000MNRAS.317.1014B} {317, 1014}

\bibitem[\protect\citeauthoryear{{Brandt} et~al.,}{{Brandt}
  et~al.}{2002}]{Brandt2002}
{Brandt} W.~N.,  et~al., 2002, \mn@doi [\apjl] {10.1086/340581}, \href
  {https://ui.adsabs.harvard.edu/abs/2002ApJ...569L...5B} {569, L5}

\bibitem[\protect\citeauthoryear{{Brunner} et~al.,}{{Brunner}
  et~al.}{2022}]{Brunner2021}
{Brunner} H.,  et~al., 2022, \mn@doi [\aap] {10.1051/0004-6361/202141266},
  \href {https://ui.adsabs.harvard.edu/abs/2022A&A...661A...1B} {661, A1}

\bibitem[\protect\citeauthoryear{{Brusa} et~al.,}{{Brusa}
  et~al.}{2009}]{Brusa2009}
{Brusa} M.,  et~al., 2009, \mn@doi [\apj] {10.1088/0004-637X/693/1/8}, \href
  {https://ui.adsabs.harvard.edu/abs/2009ApJ...693....8B} {693, 8}

\bibitem[\protect\citeauthoryear{{Burrows} et~al.,}{{Burrows}
  et~al.}{2005}]{Burrows2005}
{Burrows} D.~N.,  et~al., 2005, \mn@doi [\ssr] {10.1007/s11214-005-5097-2},
  \href {https://ui.adsabs.harvard.edu/abs/2005SSRv..120..165B} {120, 165}

\bibitem[\protect\citeauthoryear{{Carnall} et~al.,}{{Carnall}
  et~al.}{2015}]{Carnall2015}
{Carnall} A.~C.,  et~al., 2015, \mn@doi [\mnras] {10.1093/mnrasl/slv057}, \href
  {https://ui.adsabs.harvard.edu/abs/2015MNRAS.451L..16C} {451, L16}

\bibitem[\protect\citeauthoryear{{Chen}, {Pan}, {Pang}  \& {Huang}}{{Chen}
  et~al.}{2018}]{Chen2018}
{Chen} Z.-F.,  {Pan} D.-S.,  {Pang} T.-T.,   {Huang} Y.,  2018, \mn@doi [\apjs]
  {10.3847/1538-4365/aa9d90}, \href
  {https://ui.adsabs.harvard.edu/abs/2018ApJS..234...16C} {234, 16}

\bibitem[\protect\citeauthoryear{{De Rosa} et~al.,}{{De Rosa}
  et~al.}{2014}]{DeRosa2014}
{De Rosa} G.,  et~al., 2014, \mn@doi [\apj] {10.1088/0004-637X/790/2/145},
  \href {https://ui.adsabs.harvard.edu/abs/2014ApJ...790..145D} {790, 145}

\bibitem[\protect\citeauthoryear{{Delaney}, {Aird}, {Evans}, {Barlow-Hall}  \&
  {Watson}}{{Delaney} et~al.}{2022}]{DelaneyPrep}
{Delaney} J.,  {Aird} J.,  {Evans} P.~A.,  {Barlow-Hall} C.,   {Watson} M.~G.,
  2022, The Extragalactic Serendipitous Swift Survey (ExSeSS) – I. Survey
  definition and measurements of the X-ray number counts, accepted for
  publication in MNRAS, November 2022

\bibitem[\protect\citeauthoryear{{Evans} et~al.,}{{Evans}
  et~al.}{2020}]{Evans2020}
{Evans} P.~A.,  et~al., 2020, \mn@doi [\apjs] {10.3847/1538-4365/ab7db9}, \href
  {https://ui.adsabs.harvard.edu/abs/2020ApJS..247...54E} {247, 54}

\bibitem[\protect\citeauthoryear{{Gehrels}}{{Gehrels}}{1986}]{Gehrels1986}
{Gehrels} N.,  1986, \mn@doi [\apj] {10.1086/164079}, \href
  {https://ui.adsabs.harvard.edu/abs/1986ApJ...303..336G} {303, 336}

\bibitem[\protect\citeauthoryear{{Georgakakis} et~al.,}{{Georgakakis}
  et~al.}{2015}]{Georgakakis2015}
{Georgakakis} A.,  et~al., 2015, \mn@doi [\mnras] {10.1093/mnras/stv1703},
  \href {https://ui.adsabs.harvard.edu/abs/2015MNRAS.453.1946G} {453, 1946}

\bibitem[\protect\citeauthoryear{{Gilli}, {Comastri}  \& {Hasinger}}{{Gilli}
  et~al.}{2007}]{Gilli2007}
{Gilli} R.,  {Comastri} A.,   {Hasinger} G.,  2007, \mn@doi [\aap]
  {10.1051/0004-6361:20066334}, \href
  {https://ui.adsabs.harvard.edu/abs/2007A&A...463...79G} {463, 79}

\bibitem[\protect\citeauthoryear{{HI4PI Collaboration} et~al.,}{{HI4PI
  Collaboration} et~al.}{2016}]{HI4PIcollab2016}
{HI4PI Collaboration} et~al., 2016, \mn@doi [\aap]
  {10.1051/0004-6361/201629178}, \href
  {https://ui.adsabs.harvard.edu/abs/2016A&A...594A.116H} {594, A116}

\bibitem[\protect\citeauthoryear{{Hale} et~al.,}{{Hale}
  et~al.}{2021}]{Hale2021}
{Hale} C.~L.,  et~al., 2021, \mn@doi [\pasa] {10.1017/pasa.2021.47}, \href
  {https://ui.adsabs.harvard.edu/abs/2021PASA...38...58H} {38, e058}

\bibitem[\protect\citeauthoryear{{Hickox} \& {Alexander}}{{Hickox} \&
  {Alexander}}{2018}]{Hickox&Alexander2018}
{Hickox} R.~C.,  {Alexander} D.~M.,  2018, \mn@doi [\araa]
  {10.1146/annurev-astro-081817-051803}, \href
  {https://ui.adsabs.harvard.edu/abs/2018ARA&A..56..625H} {56, 625}

\bibitem[\protect\citeauthoryear{{Hopkins}, {Richards}  \&
  {Hernquist}}{{Hopkins} et~al.}{2007}]{Hopkins2007}
{Hopkins} P.~F.,  {Richards} G.~T.,   {Hernquist} L.,  2007, \mn@doi [\apj]
  {10.1086/509629}, \href
  {https://ui.adsabs.harvard.edu/abs/2007ApJ...654..731H} {654, 731}

\bibitem[\protect\citeauthoryear{Hunter}{Hunter}{2007}]{Matplotlib}
Hunter J.~D.,  2007, \mn@doi [Computing in Science Engineering]
  {10.1109/MCSE.2007.55}, 9, 90

\bibitem[\protect\citeauthoryear{{Kalfountzou}, {Civano}, {Elvis}, {Trichas}
  \& {Green}}{{Kalfountzou} et~al.}{2014}]{Kalfountzou2014}
{Kalfountzou} E.,  {Civano} F.,  {Elvis} M.,  {Trichas} M.,   {Green} P.,
  2014, \mn@doi [\mnras] {10.1093/mnras/stu1745}, \href
  {https://ui.adsabs.harvard.edu/abs/2014MNRAS.445.1430K} {445, 1430}

\bibitem[\protect\citeauthoryear{{Khorunzhev} et~al.,}{{Khorunzhev}
  et~al.}{2021}]{Khorunzhev2021}
{Khorunzhev} G.~A.,  et~al., 2021, \mn@doi [Astronomy Letters]
  {10.1134/S1063773721030026}, \href
  {https://ui.adsabs.harvard.edu/abs/2021AstL...47..123K} {47, 123}

\bibitem[\protect\citeauthoryear{{Kormendy} \& {Ho}}{{Kormendy} \&
  {Ho}}{2013}]{Kormendy&Ho2013}
{Kormendy} J.,  {Ho} L.~C.,  2013, \mn@doi [\araa]
  {10.1146/annurev-astro-082708-101811}, \href
  {https://ui.adsabs.harvard.edu/abs/2013ARA&A..51..511K} {51, 511}

\bibitem[\protect\citeauthoryear{{Luo} et~al.,}{{Luo} et~al.}{2017}]{Luo2017}
{Luo} B.,  et~al., 2017, \mn@doi [\apjs] {10.3847/1538-4365/228/1/2}, \href
  {https://ui.adsabs.harvard.edu/abs/2017ApJS..228....2L} {228, 2}

\bibitem[\protect\citeauthoryear{{Machalski}}{{Machalski}}{1998}]{Machalski1998}
{Machalski} J.,  1998, \mn@doi [\aaps] {10.1051/aas:1998132}, \href
  {https://ui.adsabs.harvard.edu/abs/1998A&AS..128..153M} {128, 153}

\bibitem[\protect\citeauthoryear{{Madau} \& {Rees}}{{Madau} \&
  {Rees}}{2001}]{Madau&Rees2001}
{Madau} P.,  {Rees} M.~J.,  2001, \mn@doi [\apjl] {10.1086/319848}, \href
  {https://ui.adsabs.harvard.edu/abs/2001ApJ...551L..27M} {551, L27}

\bibitem[\protect\citeauthoryear{{Marchesi} et~al.,}{{Marchesi}
  et~al.}{2016}]{Marchesi2016}
{Marchesi} S.,  et~al., 2016, \mn@doi [\apj] {10.3847/0004-637X/827/2/150},
  \href {https://ui.adsabs.harvard.edu/abs/2016ApJ...827..150M} {827, 150}

\bibitem[\protect\citeauthoryear{{Matsuoka} et~al.,}{{Matsuoka}
  et~al.}{2019}]{Matsuoka2019}
{Matsuoka} Y.,  et~al., 2019, \mn@doi [\apj] {10.3847/1538-4357/ab3c60}, \href
  {https://ui.adsabs.harvard.edu/abs/2019ApJ...883..183M} {883, 183}

\bibitem[\protect\citeauthoryear{{McGreer} et~al.,}{{McGreer}
  et~al.}{2013}]{McGreer2013}
{McGreer} I.~D.,  et~al., 2013, \mn@doi [\apj] {10.1088/0004-637X/768/2/105},
  \href {https://ui.adsabs.harvard.edu/abs/2013ApJ...768..105M} {768, 105}

\bibitem[\protect\citeauthoryear{{Medvedev} et~al.,}{{Medvedev}
  et~al.}{2020}]{Medvedev2020}
{Medvedev} P.,  et~al., 2020, \mn@doi [\mnras] {10.1093/mnras/staa2051}, \href
  {https://ui.adsabs.harvard.edu/abs/2020MNRAS.497.1842M} {497, 1842}

\bibitem[\protect\citeauthoryear{{Miyaji}, {Hasinger}  \& {Schmidt}}{{Miyaji}
  et~al.}{2001}]{Miyaji2001}
{Miyaji} T.,  {Hasinger} G.,   {Schmidt} M.,  2001, \mn@doi [\aap]
  {10.1051/0004-6361:20010102}, \href
  {https://ui.adsabs.harvard.edu/abs/2001A&A...369...49M} {369, 49}

\bibitem[\protect\citeauthoryear{{Mortlock} et~al.,}{{Mortlock}
  et~al.}{2011}]{Mortlock2011}
{Mortlock} D.~J.,  et~al., 2011, \mn@doi [\nat] {10.1038/nature10159}, \href
  {https://ui.adsabs.harvard.edu/abs/2011Natur.474..616M} {474, 616}

\bibitem[\protect\citeauthoryear{{Mukai}}{{Mukai}}{1993}]{PIMMS}
{Mukai} K.,  1993, Legacy, \href
  {https://ui.adsabs.harvard.edu/abs/1993Legac...3...21M} {3, 21}

\bibitem[\protect\citeauthoryear{{Nanni}, {Vignali}, {Gilli}, {Moretti}  \&
  {Brandt}}{{Nanni} et~al.}{2017}]{Nanni2017}
{Nanni} R.,  {Vignali} C.,  {Gilli} R.,  {Moretti} A.,   {Brandt} W.~N.,  2017,
  \mn@doi [\aap] {10.1051/0004-6361/201730484}, \href
  {https://ui.adsabs.harvard.edu/abs/2017A&A...603A.128N} {603, A128}

\bibitem[\protect\citeauthoryear{{Onoue} et~al.,}{{Onoue}
  et~al.}{2019}]{Onoue2019}
{Onoue} M.,  et~al., 2019, \mn@doi [\apj] {10.3847/1538-4357/ab29e9}, \href
  {https://ui.adsabs.harvard.edu/abs/2019ApJ...880...77O} {880, 77}

\bibitem[\protect\citeauthoryear{{Padovani} et~al.,}{{Padovani}
  et~al.}{2017}]{Padovani2017}
{Padovani} P.,  et~al., 2017, \mn@doi [\aapr] {10.1007/s00159-017-0102-9},
  \href {https://ui.adsabs.harvard.edu/abs/2017A&ARv..25....2P} {25, 2}

\bibitem[\protect\citeauthoryear{{Page} et~al.,}{{Page}
  et~al.}{1996}]{Page1996}
{Page} M.~J.,  et~al., 1996, \mn@doi [\mnras] {10.1093/mnras/281.2.579}, \href
  {https://ui.adsabs.harvard.edu/abs/1996MNRAS.281..579P} {281, 579}

\bibitem[\protect\citeauthoryear{{P{\^a}ris} et~al.,}{{P{\^a}ris}
  et~al.}{2017}]{Paris2017}
{P{\^a}ris} I.,  et~al., 2017, \mn@doi [\aap] {10.1051/0004-6361/201527999},
  \href {https://ui.adsabs.harvard.edu/abs/2017A&A...597A..79P} {597, A79}

\bibitem[\protect\citeauthoryear{{P{\^a}ris} et~al.,}{{P{\^a}ris}
  et~al.}{2018}]{Paris2018}
{P{\^a}ris} I.,  et~al., 2018, \mn@doi [\aap] {10.1051/0004-6361/201732445},
  \href {https://ui.adsabs.harvard.edu/abs/2018A&A...613A..51P} {613, A51}

\bibitem[\protect\citeauthoryear{{Pons}, {McMahon}, {Banerji}  \&
  {Reed}}{{Pons} et~al.}{2020}]{Pons2020}
{Pons} E.,  {McMahon} R.~G.,  {Banerji} M.,   {Reed} S.~L.,  2020, \mn@doi
  [\mnras] {10.1093/mnras/stz3275}, \href
  {https://ui.adsabs.harvard.edu/abs/2020MNRAS.491.3884P} {491, 3884}

\bibitem[\protect\citeauthoryear{{Predehl} et~al.,}{{Predehl}
  et~al.}{2021}]{Predehl2021}
{Predehl} P.,  et~al., 2021, \mn@doi [\aap] {10.1051/0004-6361/202039313},
  \href {https://ui.adsabs.harvard.edu/abs/2021A&A...647A...1P} {647, A1}

\bibitem[\protect\citeauthoryear{{Reed} et~al.,}{{Reed}
  et~al.}{2019}]{Reed2019}
{Reed} S.~L.,  et~al., 2019, \mn@doi [\mnras] {10.1093/mnras/stz1341}, \href
  {https://ui.adsabs.harvard.edu/abs/2019MNRAS.487.1874R} {487, 1874}

\bibitem[\protect\citeauthoryear{{Reines} \& {Comastri}}{{Reines} \&
  {Comastri}}{2016}]{Reines&Comastri2016}
{Reines} A.~E.,  {Comastri} A.,  2016, \mn@doi [\pasa] {10.1017/pasa.2016.46},
  \href {https://ui.adsabs.harvard.edu/abs/2016PASA...33...54R} {33, e054}

\bibitem[\protect\citeauthoryear{{Ross} \& {Cross}}{{Ross} \&
  {Cross}}{2020}]{Ross&Cross2020}
{Ross} N.~P.,  {Cross} N. J.~G.,  2020, \mn@doi [\mnras]
  {10.1093/mnras/staa544}, \href
  {https://ui.adsabs.harvard.edu/abs/2020MNRAS.494..789R} {494, 789}

\bibitem[\protect\citeauthoryear{{Ross} et~al.,}{{Ross}
  et~al.}{2013}]{Ross2013}
{Ross} N.~P.,  et~al., 2013, \mn@doi [\apj] {10.1088/0004-637X/773/1/14}, \href
  {https://ui.adsabs.harvard.edu/abs/2013ApJ...773...14R} {773, 14}

\bibitem[\protect\citeauthoryear{{Schmidt}}{{Schmidt}}{1968}]{Schmidt1968}
{Schmidt} M.,  1968, \mn@doi [\apj] {10.1086/149446}, \href
  {https://ui.adsabs.harvard.edu/abs/1968ApJ...151..393S} {151, 393}

\bibitem[\protect\citeauthoryear{{Selsing}, {Fynbo}, {Christensen}  \&
  {Krogager}}{{Selsing} et~al.}{2016}]{Selsing2016}
{Selsing} J.,  {Fynbo} J.~P.~U.,  {Christensen} L.,   {Krogager} J.~K.,  2016,
  \mn@doi [\aap] {10.1051/0004-6361/201527096}, \href
  {https://ui.adsabs.harvard.edu/abs/2016A&A...585A..87S} {585, A87}

\bibitem[\protect\citeauthoryear{{Shen} et~al.,}{{Shen}
  et~al.}{2019}]{Shen2019}
{Shen} Y.,  et~al., 2019, \mn@doi [\apj] {10.3847/1538-4357/ab03d9}, \href
  {https://ui.adsabs.harvard.edu/abs/2019ApJ...873...35S} {873, 35}

\bibitem[\protect\citeauthoryear{{Shen}, {Hopkins}, {Faucher-Gigu{\`e}re},
  {Alexander}, {Richards}, {Ross}  \& {Hickox}}{{Shen} et~al.}{2020}]{Shen2020}
{Shen} X.,  {Hopkins} P.~F.,  {Faucher-Gigu{\`e}re} C.-A.,  {Alexander} D.~M.,
  {Richards} G.~T.,  {Ross} N.~P.,   {Hickox} R.~C.,  2020, \mn@doi [\mnras]
  {10.1093/mnras/staa1381}, \href
  {https://ui.adsabs.harvard.edu/abs/2020MNRAS.495.3252S} {495, 3252}

\bibitem[\protect\citeauthoryear{{Sunyaev} et~al.,}{{Sunyaev}
  et~al.}{2021}]{Sunyaev2021}
{Sunyaev} R.,  et~al., 2021, \mn@doi [\aap] {10.1051/0004-6361/202141179},
  \href {https://ui.adsabs.harvard.edu/abs/2021A&A...656A.132S} {656, A132}

\bibitem[\protect\citeauthoryear{{Tananbaum} et~al.,}{{Tananbaum}
  et~al.}{1979}]{Tananbaum1979}
{Tananbaum} H.,  et~al., 1979, \mn@doi [\apjl] {10.1086/183100}, \href
  {https://ui.adsabs.harvard.edu/abs/1979ApJ...234L...9T} {234, L9}

\bibitem[\protect\citeauthoryear{{Taylor}}{{Taylor}}{2013}]{TopCat2013}
{Taylor} M.,  2013, Starlink User Note, \href
  {https://ui.adsabs.harvard.edu/abs/2013StaUN.253.....T} {253}

\bibitem[\protect\citeauthoryear{{Ueda}, {Akiyama}, {Hasinger}, {Miyaji}  \&
  {Watson}}{{Ueda} et~al.}{2014}]{Ueda2014}
{Ueda} Y.,  {Akiyama} M.,  {Hasinger} G.,  {Miyaji} T.,   {Watson} M.~G.,
  2014, \mn@doi [\apj] {10.1088/0004-637X/786/2/104}, \href
  {https://ui.adsabs.harvard.edu/abs/2014ApJ...786..104U} {786, 104}

\bibitem[\protect\citeauthoryear{{Vignali}, {Brandt}, {Fan}, {Gunn}, {Kaspi},
  {Schneider}  \& {Strauss}}{{Vignali} et~al.}{2001}]{Vignali2001}
{Vignali} C.,  {Brandt} W.~N.,  {Fan} X.,  {Gunn} J.~E.,  {Kaspi} S.,
  {Schneider} D.~P.,   {Strauss} M.~A.,  2001, \mn@doi [\aj] {10.1086/323712},
  \href {https://ui.adsabs.harvard.edu/abs/2001AJ....122.2143V} {122, 2143}

\bibitem[\protect\citeauthoryear{{Virtanen} et~al.,}{{Virtanen}
  et~al.}{2020}]{SciPy2020}
{Virtanen} P.,  et~al., 2020, \mn@doi [Nature Methods]
  {10.1038/s41592-019-0686-2}, \href
  {https://ui.adsabs.harvard.edu/abs/2020NatMe..17..261V} {17, 261}

\bibitem[\protect\citeauthoryear{{Vito}, {Gilli}, {Vignali}, {Comastri},
  {Brusa}, {Cappelluti}  \& {Iwasawa}}{{Vito} et~al.}{2014}]{Vito2014}
{Vito} F.,  {Gilli} R.,  {Vignali} C.,  {Comastri} A.,  {Brusa} M.,
  {Cappelluti} N.,   {Iwasawa} K.,  2014, \mn@doi [\mnras]
  {10.1093/mnras/stu2004}, \href
  {https://ui.adsabs.harvard.edu/abs/2014MNRAS.445.3557V} {445, 3557}

\bibitem[\protect\citeauthoryear{{Vito} et~al.,}{{Vito}
  et~al.}{2019}]{Vito2019}
{Vito} F.,  et~al., 2019, \mn@doi [\aap] {10.1051/0004-6361/201936217}, \href
  {https://ui.adsabs.harvard.edu/abs/2019A&A...630A.118V} {630, A118}

\bibitem[\protect\citeauthoryear{{Volonteri} \& {Begelman}}{{Volonteri} \&
  {Begelman}}{2010}]{Volonteri2010}
{Volonteri} M.,  {Begelman} M.~C.,  2010, \mn@doi [\mnras]
  {10.1111/j.1365-2966.2010.17359.x}, \href
  {https://ui.adsabs.harvard.edu/abs/2010MNRAS.409.1022V} {409, 1022}

\bibitem[\protect\citeauthoryear{{Wang} et~al.,}{{Wang}
  et~al.}{2019}]{Wang2019}
{Wang} F.,  et~al., 2019, \mn@doi [\apj] {10.3847/1538-4357/ab2be5}, \href
  {https://ui.adsabs.harvard.edu/abs/2019ApJ...884...30W} {884, 30}

\bibitem[\protect\citeauthoryear{{Wang} et~al.,}{{Wang}
  et~al.}{2021}]{Wang2021}
{Wang} F.,  et~al., 2021, \mn@doi [\apjl] {10.3847/2041-8213/abd8c6}, \href
  {https://ui.adsabs.harvard.edu/abs/2021ApJ...907L...1W} {907, L1}

\bibitem[\protect\citeauthoryear{{Wenger} et~al.,}{{Wenger}
  et~al.}{2000}]{SimbadDatabase}
{Wenger} M.,  et~al., 2000, \mn@doi [\aaps] {10.1051/aas:2000332}, \href
  {https://ui.adsabs.harvard.edu/abs/2000A&AS..143....9W} {143, 9}

\bibitem[\protect\citeauthoryear{{Wolf} et~al.,}{{Wolf}
  et~al.}{2021}]{Wolf2021}
{Wolf} J.,  et~al., 2021, \mn@doi [\aap] {10.1051/0004-6361/202039724}, \href
  {https://ui.adsabs.harvard.edu/abs/2021A&A...647A...5W} {647, A5}

\bibitem[\protect\citeauthoryear{{Wolf} et~al.,}{{Wolf}
  et~al.}{2022}]{Wolf2022}
{Wolf} J.,  et~al., 2022, arXiv e-prints, \href
  {https://ui.adsabs.harvard.edu/abs/2022arXiv221113820W} {p. arXiv:2211.13820}

\bibitem[\protect\citeauthoryear{{Yang} et~al.,}{{Yang}
  et~al.}{2021}]{Yang2021}
{Yang} J.,  et~al., 2021, \mn@doi [\apj] {10.3847/1538-4357/ac2b32}, \href
  {https://ui.adsabs.harvard.edu/abs/2021ApJ...923..262Y} {923, 262}

\bibitem[\protect\citeauthoryear{{Zubovas} \& {King}}{{Zubovas} \&
  {King}}{2021}]{Zubovas&King2021}
{Zubovas} K.,  {King} A.,  2021, \mn@doi [\mnras] {10.1093/mnras/stab004},
  \href {https://ui.adsabs.harvard.edu/abs/2021MNRAS.501.4289Z} {501, 4289}

\bibitem[\protect\citeauthoryear{{van der Walt}, {Colbert}  \&
  {Varoquaux}}{{van der Walt} et~al.}{2011}]{NumPy}
{van der Walt} S.,  {Colbert} S.~C.,   {Varoquaux} G.,  2011, \mn@doi
  [Computing in Science and Engineering] {10.1109/MCSE.2011.37}, \href
  {https://ui.adsabs.harvard.edu/abs/2011CSE....13b..22V} {13, 22}

\makeatother
\end{thebibliography}
\end{document}